\documentclass[12pt]{article}

\usepackage{scicite}
\usepackage{xcolor}
\usepackage{graphicx}

\usepackage{amsmath,siunitx}
\usepackage{amssymb}
\usepackage{physics}
\usepackage{hyperref}
\usepackage[normalem]{ulem}

\def \arcsec {\hbox{$^{\prime\prime}$}}

\usepackage{times}

\newenvironment{sciabstract}{%
\begin{quote} \bf}
{\end{quote}}

\newcounter{lastnote}

\title{A structured jet explains the extreme GRB 221009A}

\author{
B. O'Connor$^{1,2,3,4\ast}$, 
E. Troja$^{5,6\ast}$, 
G. Ryan$^{7}$,
P. Beniamini$^{8,9}$,
H. van Eerten$^{10}$, \\
J. Granot$^{8,9,1}$, 
S. Dichiara$^{11}$,
R. Ricci$^{12,13}$,
V. Lipunov$^{14}$, 
J. H. Gillanders$^{5}$, 
R. Gill$^{15}$, \\
M. Moss$^{1}$, 
S. Anand$^{16}$, 
I. Andreoni$^{17,3,4,\dagger}$, 
R. L. Becerra$^{18}$, 
D. A. H. Buckley$^{19,20}$,\\
N. R. Butler$^{21}$, 
S. B. Cenko$^{4,17}$, 
 A. Chasovnikov$^{14}$, 
J. Durbak$^{3,4}$, 
C. Francile$^{22,23}$,   \\
E. Hammerstein$^{3}$,
A. J. van der Horst$^{1}$,
M. Kasliwal$^{16}$, 
C. Kouveliotou$^{1,2}$, \\
A. S. Kutyrev$^{3,4}$,  
W. H. Lee$^{24}$, 
G. Srinivasaragavan$^{3}$, 
V. Topolev$^{14}$, \\
A. M. Watson$^{24}$, 
Y.-H. Yang$^5$, 
K. Zhirkov$^{14}$
}

\date{}

\begin{document} 


\baselineskip24pt


\maketitle 

\begin{center}
    \scriptsize
    \baselineskip24pt
    {$^{1}$ Department of Physics, The George Washington University, 725 21st Street NW, Washington, DC 20052, USA } \\ 
    {$^{2}$ Astronomy, Physics and Statistics Institute of Sciences (APSIS), Washington, DC 20052, USA }\\
    {$^{3}$ Department of Astronomy, University of Maryland, College Park, MD 20742-4111, USA }\\
    {$^{4}$ Astrophysics Science Division, NASA Goddard Space Flight Center, 8800 Greenbelt Rd, Greenbelt, MD 20771, USA }\\
    {$^{5}$ Department of Physics, University of Rome ``Tor Vergata'', via della Ricerca Scientifica 1, I-00133 Rome, Italy}\\
    {$^{6}$ INAF - Istituto di Astrofisica e Planetologia Spaziali, via Fosso del Cavaliere 100, 00133 Rome, Italy} \\
    {$^{7}$ Perimeter Institute for Theoretical Physics, 31 Caroline St. N., Waterloo, ON, N2L 2Y5, Canada }\\
    {$^{8}$ Department of Natural Sciences, The Open University of Israel, P.O Box 808, Ra’anana 4353701, Israel }\\
    {$^{9}$ Astrophysics Research Center of the Open university (ARCO), P.O Box 808, Ra’anana 4353701, Israel }\\
    {$^{10}$ Physics Department, University of Bath, Claverton Down, Bath BA2 7AY, United Kingdom}\\
    {$^{11}$ The Pennsylvania State University, 525 Davey Lab, University Park, PA 16802, USA} \\
    {$^{12}$ Istituto Nazionale di Ricerche Metrologiche, I-10135 Torino, Italy}\\
    {$^{13}$ INAF - Istituto di Radioastronomia, via Gobetti 101, I-40129 Bologna, Italy}\\
    {$^{14}$ Sternberg Astronomical Institute, Lomonosov Moscow State University, 119234, Universitetsky, 13, Moscow, Russia} \\
    {$^{15}$ Instituto de Radioastronom\'ia y Astrof\'isica, Universidad Nacional Aut\'onoma de M\'exico, Antigua Carretera } \\ 
    {a P\'atzcuaro $\#$ 8701, Ex-Hda. San Jos\'e de la Huerta, Morelia, Michoac\'an, C.P. 58089, M\'exico } \\
    {$^{16}$ Division of Physics, Mathematics and Astronomy, California Institute of Technology, Pasadena, CA 91125, USA}\\
    {$^{17}$ Joint Space-Science Institute, University of Maryland, College Park, MD 20742, USA}\\
    {$^{18}$ Instituto de Ciencias Nucleares, Universidad Nacional Aut\'onoma de M\'exico, 
    04510 M\'exico, CDMX, Mexico}\\
    {$^{19}$ Department of Astronomy, University of Cape Town, Private Bag X3, Rondebosch 7701, South Africa}\\
    {$^{20}$ South African Astronomical Observatory, PO Box 9, 7935 Observatory, Cape Town, South Africa}\\
    {$^{21}$ School of Earth and Space Exploration, Arizona State University, Tempe, AZ 85287, USA }\\
    {$^{22}$ Observatorio Astronomico Felix Aguilar (OAFA), 
    San Juan 5400, Argentina} \\
    {$^{23}$ Facultad de Ciencias Exactas Fisicas y Naturales, San Juan National University, San Juan 5400, Argentina}\\
    {$^{24}$ Instituto de Astronom\'ia, Universidad Nacional Aut\'onoma de M\'exico, 04510 M\'exico, CDMX, Mexico }\\
    {$^\dagger$ Neil Gehrels Fellow}\\
    {$^\ast$Correspondence to: oconnorb@gwmail.gwu.edu, eleonora.troja@uniroma2.it}
\end{center}

\paragraph*{One-sentence Summary:} 
The afterglow of GRB 221009A implies a shallow structured jet is produced in the most powerful stellar explosions.

\begin{sciabstract}
Long duration gamma-ray bursts (GRBs) are powerful cosmic explosions, signaling the death of massive stars. Among them, GRB 221009A is by far the brightest burst ever observed. Due to its enormous energy ($E_\textrm{iso}\!\approx$10$^{55}$ erg) and proximity ($z\!\approx$0.15), GRB 221009A is an exceptionally rare event that pushes the limits of our theories. We present multi-wavelength observations covering the first three months of its afterglow evolution. The X-ray brightness decays as a power-law with slope $\approx$\,$t^{-1.66}$, which is not consistent with standard predictions for jetted emission. We attribute this behavior to a shallow energy profile of the relativistic jet. A similar trend is observed in other energetic GRBs, suggesting that the most extreme explosions may be powered by structured jets launched by a common central engine.
\end{sciabstract}

\section*{Main Text}


Gamma-ray bursts (GRBs) are sudden and brief flashes of high energy radiation. Those lasting longer than a couple of seconds generally signal the death of very massive, rapidly rotating stars.
With typical durations of 1\,$-$\,100 s \cite{vonKienlin2020}, and isotropic-equivalent luminosities of $10^{50}-10^{54}$~erg~s$^{-1}$ \cite{Salvaterra2012}, they are considered one of the most energetic explosions in the Universe. When their intense radiation reaches us, it is attenuated by the large distance scale it has travelled, 
$\approx$\,16~Gpc for the median GRB redshift $z$\,$\sim$\,2
\cite{Hjorth2012}. Moreover, most of the flux above gigaelectronvolt (GeV) energies is suppressed by interactions with the extragalactic background light (EBL; \cite{Franceschini2008, Dominguez2011}). 
Therefore, as observed at Earth, GRBs display fluences in the range $10^{-7}$\,$-$\,$10^{-4}$ erg cm$^{-2}$ \cite{vonKienlin2020}
and spectra up to the MeV or, less frequently, GeV range \cite{Ajello2019}. 

\begin{figure*} 
\centering
\includegraphics[width=\textwidth]{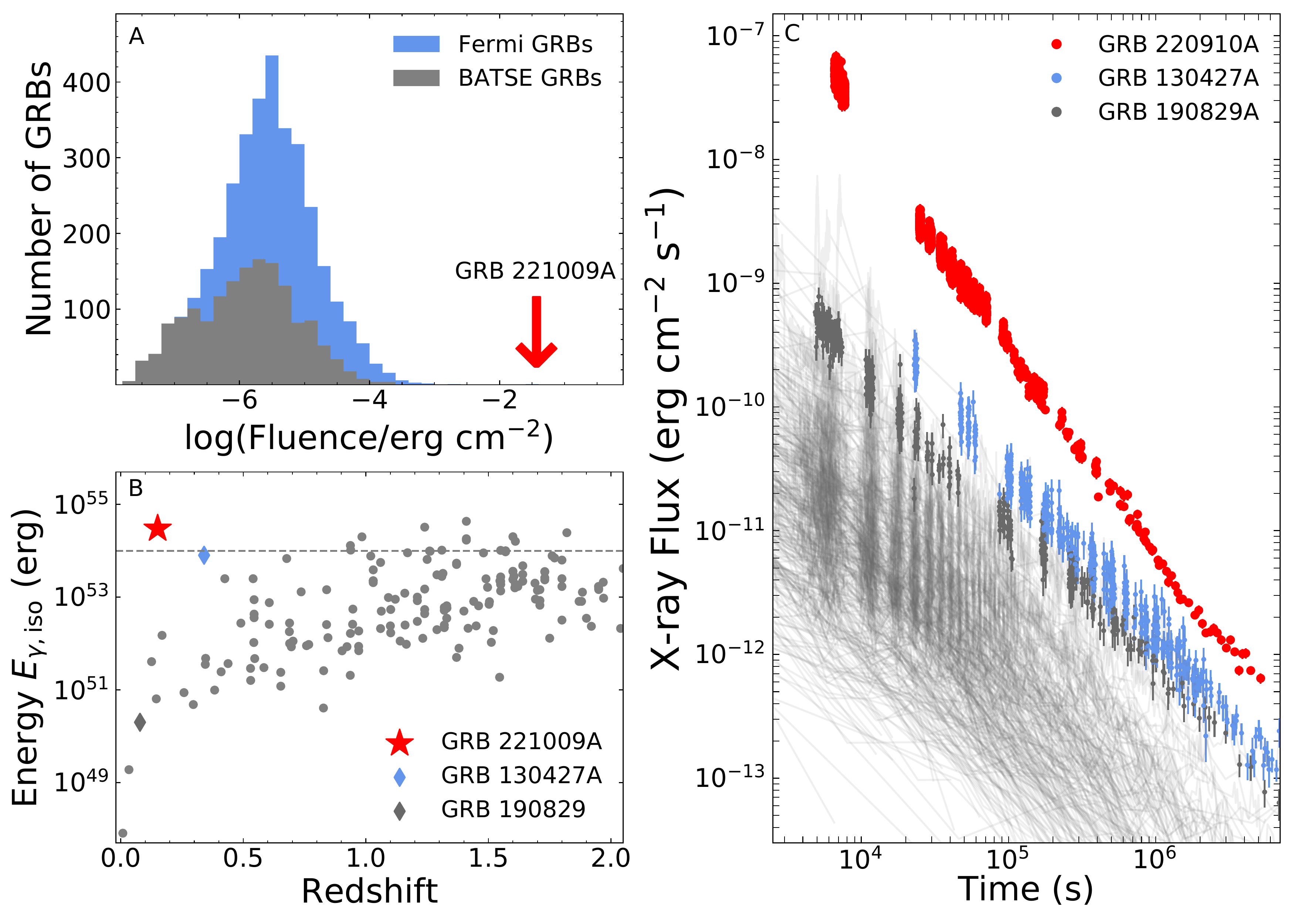}
\caption{\textbf{The extreme brightness of GRB 221009A. }
Panel A (top left): Histogram of gamma-ray fluence for \textit{Fermi} (blue; $10-1000$ keV) and \textit{BATSE} (gray; $50-300$ keV) GRBs compared to GRB 221009A. 
Panel B (bottom left): Isotropic-equivalent gamma-ray energy \mbox{(1 keV - 10 MeV)} versus redshift for a sample of long-duration GRBs compiled from \cite{Atteia2017,Lien2016}. 
Panel C (right): Observed X-ray afterglow lightcurves in the \textit{Swift} XRT ($0.3-10$ keV) energy band for a sample of long-duration GRBs. GRB 221009A is the brightest X-ray afterglow ever observed.
}
\label{fig: bright_xrt_lightcurve}
\end{figure*}

On October 10, 2022 at 13:16:59 UT (hereafter referred to as $T_0$), the Gamma-ray Burst Monitor (GBM) aboard \textit{Fermi} \cite{GCN32636,GCN32642}, among many other high-energy satellites (\textit{INTEGRAL}, \textit{Konus-Wind}, \textit{AGILE}, \textit{SRG}, \textit{GRBAlpha}, \textit{HEBS}; \cite{GCN32660,GCN32663,GCN32668, GCN32685, GCN32650}), detected an unprecedented, extremely bright burst lasting hundreds of seconds. 
This burst, dubbed GRB 221009A, is the brightest GRB ever detected in nearly 55 years of operating gamma-ray observatories, with an observed fluence of $\approx$\,$5\times 10^{-2}$ erg cm$^{-2}$ in the 20 keV\,$-$\,10 MeV band, more than an order of magnitude brighter than 
GRB 840304 and GRB 130427A \cite{Fermi130427A}, the previous record holders (Fig. \ref{fig: bright_xrt_lightcurve}). Its high-energy radiation was so intense that it disturbed Earth's ionosphere \cite{GCN32744,Hayes2022}. 

The prompt gamma-ray phase was followed by longer-lived, non-thermal afterglow radiation, visible across nearly 19 decades in energy, from low-frequency radio up to the teraelectronvolt range, corresponding to the highest energy photon (18 TeV) ever detected from a GRB \cite{GCN32677}. The afterglow phase was exceptionally bright at all frequencies and at all times, surpassing the population of X-ray afterglows by over an order of magnitude (Fig. \ref{fig: bright_xrt_lightcurve}) and causing the \textit{Neil Gehrels Swift Observatory} 
to send a trigger alert nearly an hour ($T_0$\,$+$\,$55$ min) after the initial gamma-rays were detected for the first time \cite{Williams2023}.

The extraordinary properties of this GRB are only partially explained by its proximity to us. At a redshift of $z$\,$=$\,$0.1505$ \cite{GCN32648xshooterz}, %
its luminosity distance is $\approx$\,720 Mpc \cite{Freedman2021}, 
a factor of $\gtrsim$\,20 closer than the average GRB. 
However, even after correcting for distance effects, GRB 221009A remains one of the most luminous explosions to date,  pushing the limits of our understanding in terms of both GRB energetics and the rate of events. 
Its isotropic-equivalent gamma-ray energy, $E_{\gamma,\textrm{iso}}$\,$\gtrsim$\,$3\times 10^{54}$ erg measured over the 20 keV\,$-$\,10 MeV energy range \cite{GCN32668}, 
is at the top of the GRB energy distribution (Fig. \ref{fig: bright_xrt_lightcurve}) and only sets a lower limit to the total (isotropic-equivalent) energy release. 
By including the blastwave kinetic energy that is converted into afterglow radiation, as well as the contribution of the TeV component, the isotropic-equivalent energy budget would easily surpass
10$^{55}$ erg, corresponding to $\gtrsim$\,5\,$M_{\odot}c^2$.
According to the GRB luminosity function \cite{Salvaterra2012},
an event as bright as GRB 221009A occurs this close to Earth 
less than once in a century. 
If we factor in its long duration and total energy release, our chance to observe a similar event is 1 in $\approx$\,1,000 yrs (see materials and methods). 
The detection of GRB~221009A and other extraordinary events, such as GRB 130427A \cite{Fermi130427A}, seems therefore at odds with our basic expectations of how frequent the most energetic explosions are in the nearby Universe. 

A key element for calculating the true energy release and rate of events is the geometry of the relativistic outflow. 
The outflow's angular structure and collimation leave clear imprints in GRB afterglow lightcurves \cite{BGG2020,Ryan2020,vanEerten13,Rhoads1999}, and, therefore, we can constrain these properties through our multi-wavelength campaign. In particular, if the outflow is collimated into narrow sharp-edged jets, we should observe the afterglow flux rapidly falling off after the time of the `jet-break', i.e., when the inverse of the Lorentz factor of the outflow becomes comparable to the jet's half-opening angle $\theta_{\rm j}$
\cite{Rhoads1999,Sari1999}. 
To search for the signature of collimation, we turn to the X-ray afterglow, which is unaffected by other components (e.g. supernova, reverse shock) and probes the non-thermal emission from electrons accelerated by the forward shock, driven by the outflow into the surrounding medium \cite{Meszaros1997}. 

\begin{figure*} 
\centering
\includegraphics[width=0.95\textwidth]{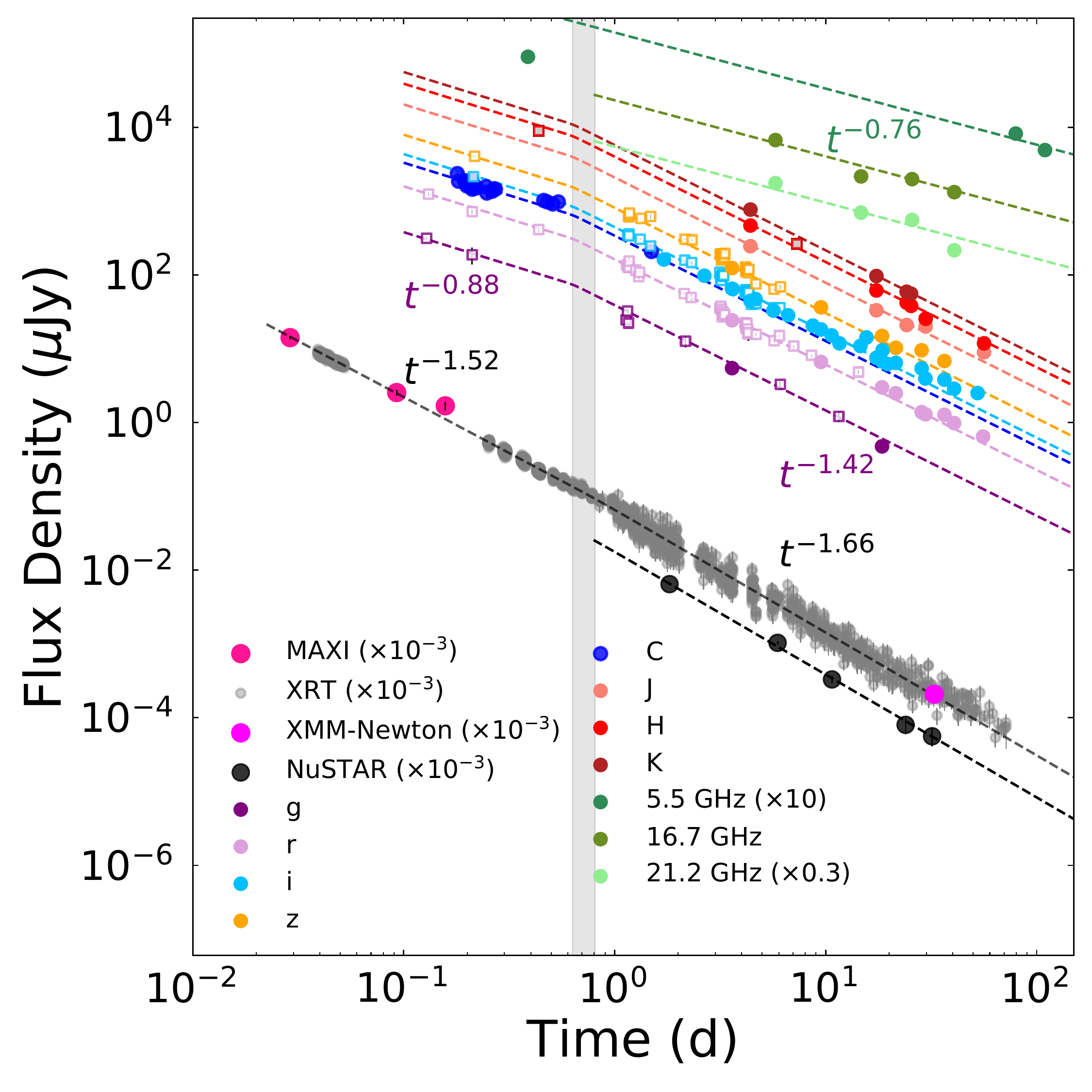}
\caption{\textbf{Multi-wavelength lightcurve of GRB 221009A.} The XRT and \textit{XMM-Newton} data represent the X-ray flux density at 1 keV, whereas the flux density from the \textit{NuSTAR} observations is reported at 5 keV. OIR data represented as empty squares were compiled from GCN circulars, whereas filled circles are data analyzed in this work. The OIR data is not corrected for Galactic extinction. }
\label{fig: multiwavelength_lightcurve}
\end{figure*}

The X-ray lightcurve features an initial power-law decay index of $\alpha_{\textrm{X},1}$\,$=$\,$-1.52\pm0.01$, steepening to $\alpha_{\textrm{X},2}$\,$=$\,$-1.66\pm0.01$ after $t_{\textrm{b},X}$\,$=$\,$0.82\pm0.07$ d (see Fig. \ref{fig: multiwavelength_lightcurve}). 
The X-ray spectrum is well described by an absorbed power-law with a time-variable spectral index, ranging from $-0.65\pm0.02$ measured by \textit{Swift} at 1 hr to $-1.10\pm0.17$ measured by \textit{NuSTAR} at 32 d. 
According to standard models of GRB jets \cite{Meszaros1997,Sari1998}, this progressive softening is consistent with the passage of the cooling frequency $\nu_c$ of the synchrotron spectrum. Therefore, the X-ray spectral shape can be used to constrain the density profile of the circumburst medium as $\rho(r) \propto r^{-k}$ with $k$\,$<$\,$4/3$ (such that $\nu_c$ decreases with time), and the energy distribution of the shock-accelerated electrons as $N(E)$\,$\propto$\,$E^{-p}$ with $p$\,
$\approx$\,$2.2$\,$-$\,$2.4$. 
This value matches the spectral measurements of the early high-energy emission \cite{GCN32658, GCN32650}, suggesting that the synchrotron component extends to the GeV range. 
Thus, we can use the high energy flux, assuming it is afterglow dominated, as a proxy for the blastwave kinetic energy \cite{Beniamini2015}, 
obtaining
$E_{\rm K,iso}$\,$\approx$\,$10^{55} \left(1+Y\right)$  erg,
where $Y$ is the Compton parameter of the GeV emitting electrons. 
A similarly high value is obtained by assuming a typical gamma-ray efficiency $\eta_\gamma$\,$\approx$\,$20\%$. 

In the standard model, the GRB jet has a constant energy profile within its core of angular size $\theta_{\rm j}$. The energy then declines rapidly or goes to zero at angles beyond $\theta_{\rm j}$. 
The prediction for the post jet-break decay is $t^{-p}$\,$\approx$\,$t^{-2.2}$, which is inconsistent with the X-ray slope of $-1.66$ measured after $t_{\textrm{b},X}$. 
If $t_{\textrm{b},X}$ is not the jet-break time $t_{\textrm{j}}$, then the uninterrupted power-law decay of the X-ray emission sets a lower limit $t_{\textrm{j}}$\,$>$\,80 d for the jet-break time. 
The resulting limit on the jet opening angle, $\theta_{\rm j} \gtrsim 15^{\circ}$, pushes the total collimation-corrected energy release to 
\mbox{$E_\textrm{K}$\,$\gtrsim$\,$4\times10^{53}$  $(t_{\textrm{j}}/{\textrm{80 d}})^{0.75}$~erg} (see materials and methods), leading to an energy crisis for most models of GRB central engines \cite{USOV1992,NARAYAN2001}. 
However, at the time of the X-ray temporal break, 
the optical and infrared (hereafter, OIR) lightcurves are also seen to steepen. The OIR emission displays an initial shallow decay with $\alpha_{\textrm{OIR},1}$\,$=$\,$-0.88\pm0.05$, which steepens to \mbox{$\alpha_{\textrm{OIR},2}$\,$=$\,$-1.42\pm0.11$} at around $t_{\textrm{b,OIR}}$\,$=$\,$0.63\pm0.13$ d (Fig. \ref{fig: multiwavelength_lightcurve}). 
The achromatic steepening of the X-ray and OIR lightcurves 
provides a strong indication of a geometrical effect, such as a jet-break \cite{Rhoads1999,Sari1999}, although the observed post-break decay rates are shallower than theoretical predictions. 
If GRB 221009A was followed by a supernova, like most long GRBs are, the supernova contribution could cause an apparent flattening of the OIR lightcurve and mask the jet-break.
By assuming that a SN~1998bw-like transient contributes to the OIR emission, an afterglow decay rate as steep as $-1.5$ is consistent with the optical and near-IR data. This is close to the observed X-ray slopes, yet too shallow for a post-jet-break phase. 

Additional evidence for geometrical effects comes from the late-time radio observations, which tend to favor a collimated outflow. In fact, the X-ray flux at 1 hr sets a lower limit of $\gtrsim$10~mJy to the forward shock peak flux. As the shock cools down and passes from the X-rays to the radio band, the peak brightness remains constant in an uniform medium \cite{Granot2002} or slowly decreases as $F_{\rm\nu,max}$\,$\propto$\,$t^{-\alpha_k}$, with $\alpha_k$\,$<$\,$1/4$
in a stratified environment with $k$\,$<$\,$4/3$. 
Either behavior would violate the observed radio limits of 0.4 mJy at 80 d unless the assumption of spherical symmetry breaks down, 
causing the peak flux to decrease more rapidly \cite{Sari1999}.

\begin{figure} 
\centering
\includegraphics[width=1.\columnwidth]{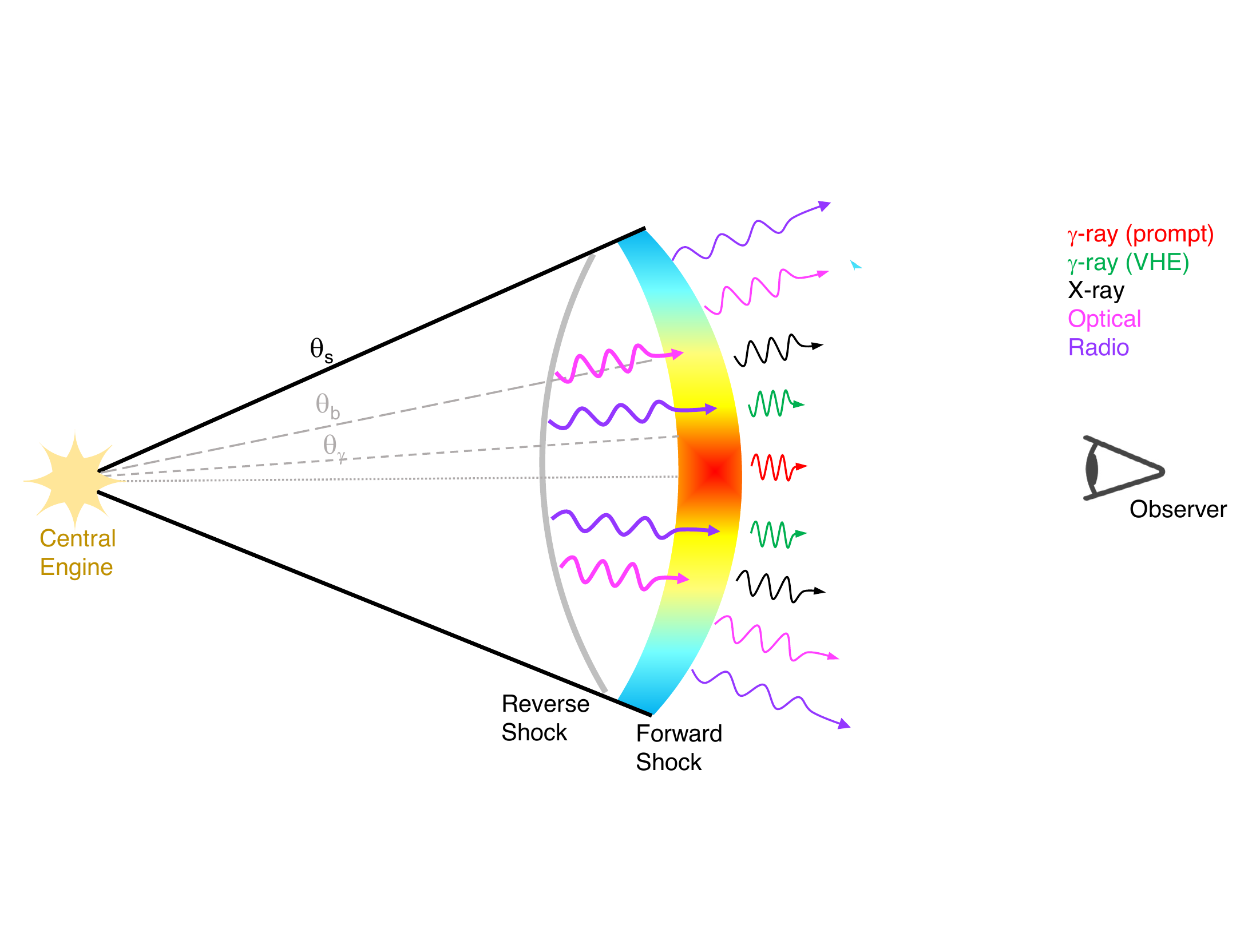}
\caption{\textbf{Schematic of the structured jet for GRB 221009A.} Emission from the forward and reverse shocks are produced by the jet out to its truncation angle $\theta_s$. The angular structure of the jet,  $dE_\textrm{K}/d\Omega$\,$\propto$\,$\theta^{-a}$, 
breaks slightly at $\theta_b$, transitioning from a slope $a_1$\,$\sim$\,0.75 to $a_2$\,$\sim$\,1.15.
The prompt gamma-rays may be radiated from the central narrow core of aperture $\theta_\gamma$. 
}
\label{fig: schematic}
\end{figure}

These different and apparently discordant observations can be reconciled if the afterglow emission is powered by a structured jet with a shallow angular energy profile \cite{BGG2022,Lamb21,Rossi04}, composed by an inner component of angular size $\theta_b$ with a shallow energy profile $dE_\textrm{K}/d\Omega$\,$\propto$\,$\theta^{-a_1}$, and a slightly steeper lateral structure at $\theta$\,$>$\,$\theta_b$
with $dE_\textrm{K}/d\Omega$\,$\propto$\,$\theta^{-a_2}$, where $a_1$\,$<$\,$a_2$\,$<$\,$2$ (see Fig. \ref{fig: schematic}).
This profile is motivated by the the lack of a sharp jet break feature in the X-ray and OIR lightcurves and the energy crisis that would be implied for a jet with a steeper angular profile. 
Similar shallow angular profiles are seen in simulations of relativistic jets expanding in complex media \cite{Gottlieb2021}.

\begin{figure*}
\centering
\includegraphics[width=0.95\textwidth]{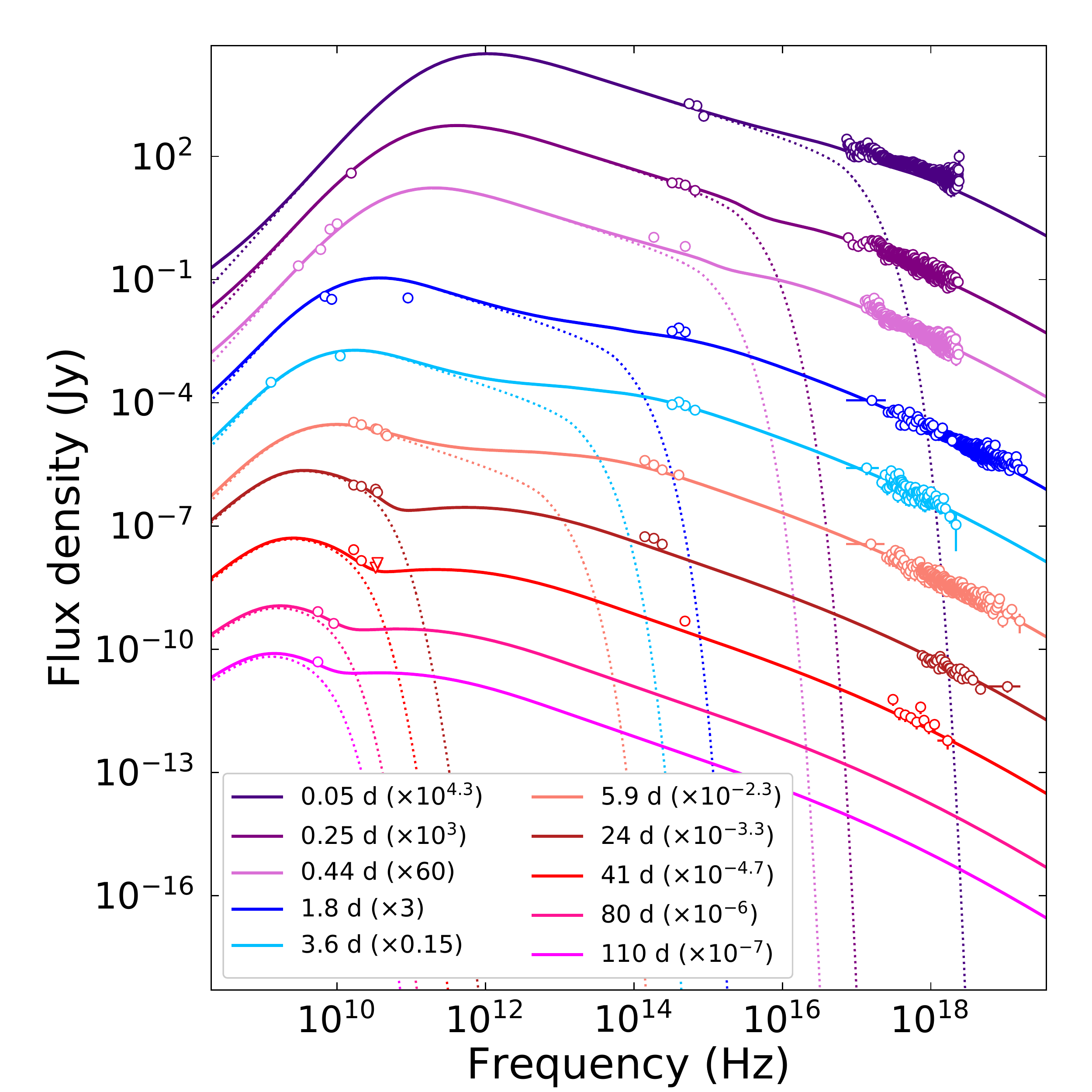}
\caption{\textbf{Afterglow spectral evolution.} Multi-epoch broad-band spectral energy distributions (SEDs) of GRB 221009A modeled by the combination (solid line) of forward shock and reverse shock (dotted line); see materials and methods. The data is corrected for extinction and absorption. 
}
\label{fig: multiwavelength_SED}
\end{figure*}

A structured jet  can account for the achromatic temporal break visible at X-ray and OIR wavelengths, and explain their post-break slopes as emission from the lateral structure as it comes into view. For $a_1$\,$\sim$\,$0.75$, $a_2$\,$\sim$\,$1.15$ and a transition at $\theta_b$\,$\sim$\,$3^{\circ}$, this model 
yields initial temporal slopes of $\alpha_\textrm{\textrm{OIR},1}$\,$=$\,$-1.3$ and $\alpha_{X,1}$\,$=$\,$-1.55$ that transition to  $\alpha_\textrm{\textrm{OIR},2}$\,$=$\,$-1.47$ and $\alpha_{X,2}$\,$=$\,$-1.67$ after $\sim$\,0.8 d. 
Although this does not capture the complexity of the early-time afterglow evolution, it provides a good description of the full X-ray lightcurve, and the OIR dataset from 0.8 d to 80 d (Fig. \ref{fig: multiwavelength_SED}). 
Over this time period, the low-frequency radio counterpart is dominated by synchrotron emission from the ejecta decelerated by the reverse shock. Emission from the reverse shock is likely contributing to the optical light curve at $t$\,$<$\,0.8 d, and responsible for its early shallow decay. The evolution of this component, however, does not follow standard prescriptions.

The main advantage of the structured jet model is that it eases up the energetic requirements relative to the uniform jet, leading to 
$E_\textrm{K}$\,$\lesssim$\,$8 \times 10^{52}$\,$(t_{\rm j}/{80\mbox{ d}})^{0.37}$~erg (see materials and methods), 
where $t_{\rm j}$ is the observed time when the jet edges become visible, causing a final steepening of the lightcurve (if still relativistic). Causal contact across the full jet surface will only be established once the jet edges are already sub-relativistic, leaving no room for strong jet spreading before the observed X-ray lightcurve behaviour segues into a non-relativistic slope $t^{(4-3p)/2}$
\cite{Frail2000, VanEerten2012} that resembles the pre-transition slope. 
In the structured jet model, the collimation-corrected energy remains at the boundary of the energy budget for a magnetar central engine ($<$\,$10^{53}$~erg for a rapidly-rotating supramassive neutron star), requiring an unrealistically high efficiency in converting the magnetar's rotational output into gamma-rays and blastwave kinetic energy. 
The massive energy required to power GRB 221009A is consistent with a magneto-hydrodynamical process \cite{BlandfordZnajek1977}, extracting
rotational energy from a rapidly spinning ($a$\,$=$\,$0.9$) stellar mass ($\sim$\,$5$\,$M_\odot$) black hole \cite{Cenko2011}.

\begin{figure*}
\centering
\includegraphics[width=0.85\textwidth]{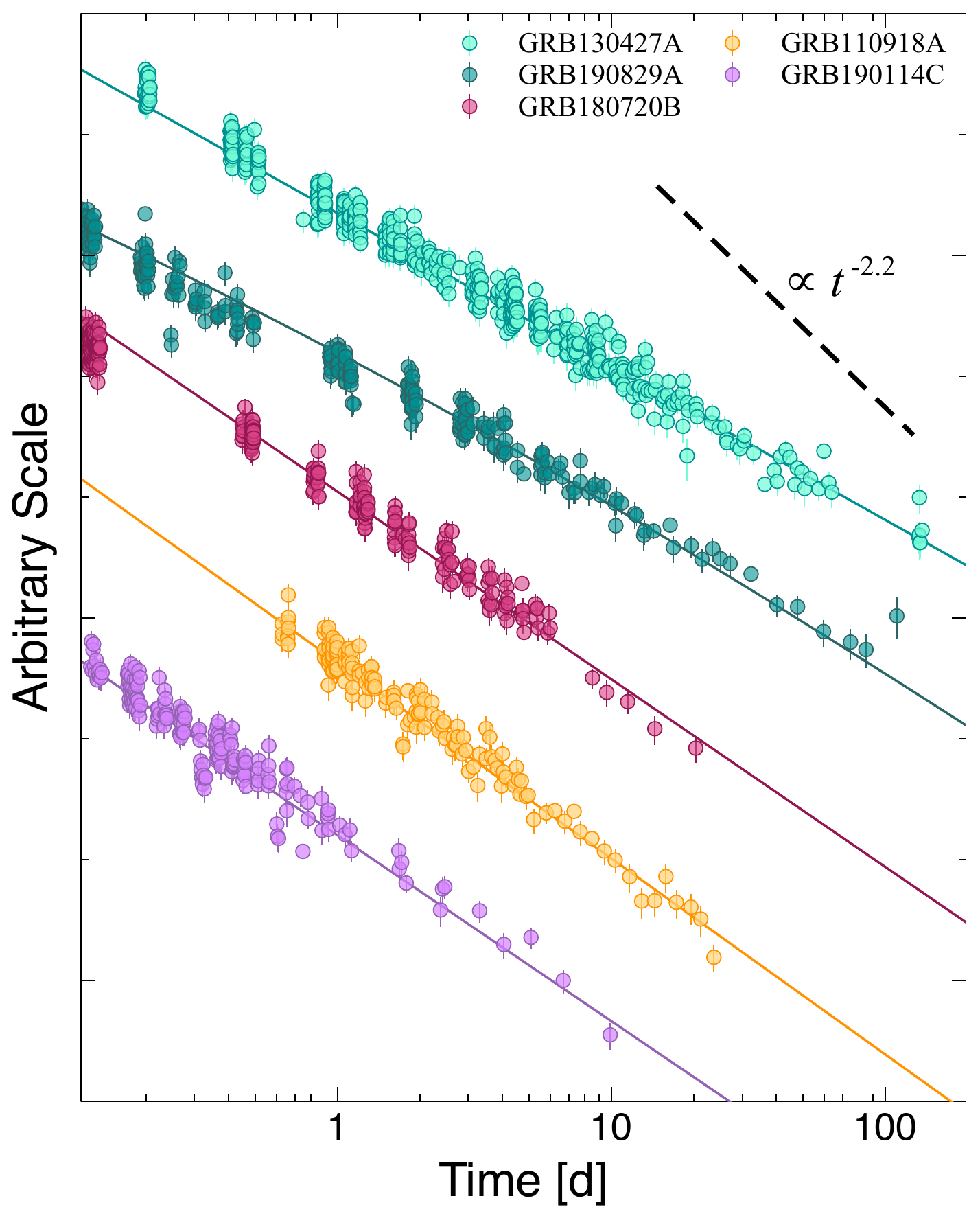}
\caption{\textbf{Long-lived X-ray lightcurves of bright GRBs.} A sample of bright long GRBs without a canonical jet-break to late-times is shown. For comparison, the dashed line shows the predicted late-time decay for a sharp-edged uniform jet. 
}
\label{fig: longgrb}
\end{figure*}

The shallow structured jet model helps explain the lack of prominent jet-breaks in some long GRBs \cite{Liang07,Racusin2009}. 
In particular, the family of bright bursts with very high-energy emission, including events like 
GRB 130427A, GRB 180720B, GRB 190114C, and GRB 190829A, 
share the common property of long-lasting afterglows (Fig. \ref{fig: longgrb}) with late-time temporal decay indices between 1.4 and 1.7 \cite{Dichiara2022,DePasquale2016,Misra2021}, similar to GRB 221009A. 
A shallow angular structure may thus be a frequent feature of the most violent explosions. 
However, none of these bursts reached the high energy of GRB 221009A, which provides compelling evidence for revising the standard jet model in a massive star explosion.

A structured jet profile also affects the rate calculation. For GRB 221009A we infer an angular size, $\theta_s$\,$\gtrsim$\,$0.4$ rad (see materials and methods), larger than that of the general GRB population, $\theta_{\rm j}$\,$\approx$\,$0.1$ rad \cite{Wang2018}. 
This would naively suggests that a similar event is $(\theta_s/\theta_{\rm j})^2\approx 16$ times more likely to be detected than a regular GRB. 
If the intrinsic rate of highly energetic GRBs is significantly lower than the rate of typical GRBs, then the larger solid angle of the jet can explain the detection of GRB 221009A.
However, this interpretation is not supported by the small viewing angle, $\theta_{\rm obs}$\,$\lesssim$\,$0.01$ rad, inferred from afterglow observations.  
Alternatively, if the rate of highly energetic GRBs
is comparable to the rate of standard GRBs, the large angular size of GRB 221009A is not consistent with the low rate of observed events (Fig. \ref{fig: bright_xrt_lightcurve}).
A natural explanation for this contradiction is that the prompt gamma-ray radiation is produced only within a narrow range of the GRB jet ($\theta_\gamma$\,$\ll$\,$\theta_s$) \cite{Berger2003},
due to a reduction in Lorentz factor $\Gamma$ with angle.
This leads to an increased opacity to photon-photon annihilation, which in turn can suppress the emission beyond a critical value \cite{Lithwick2001}, and for a wide range of dissipation and emission mechanism models a smaller Lorentz factor also leads to a reduction in the gamma-ray production. One example is a decrease in the dissipation radius $R_d$\,$\propto$\,$\Gamma^2$ at which the gamma-rays are emitted. 
Even a small reduction in Lorentz factor can lead the dissipation radius to be smaller than the photospheric radius, which decreases with Lorentz factor,
trapping the gamma-ray radiation for angles away from the core. 
This would lower the total energy released in gamma-rays by a factor ($\theta_s/\theta_\gamma$)$\gtrsim$\,20, but still require a substantial radiative efficiency along the sightline, $\eta_{\gamma}(\theta_{obs}) \gtrsim$20\%.

The suppression of gamma-ray emission above $\theta_\gamma$ would cause observers at $\theta_\textrm{obs}$\,$>$\,$\theta_\gamma$ not to detect the prompt GRB emission, and instead possibly identify such an event as an ``orphan'' afterglow \cite{Rhoads2003}. 
This may lead to a population of luminous orphan afterglows, which could be searched for in various transient surveys \cite{Cenko13}. 
The predicted rate of orphan afterglows differs by orders of magnitude between different jet models. However, as the jet angular structure shapes the early afterglow evolution \cite{Ryan2020}, 
search strategies calibrated on a uniform jet model may not efficiently recover all the possible events. 
To constrain the rate of GRB 221009A-like transients, and their gamma-ray beaming factor, 
transient classification schemes should be fine-tuned to a wide range of angular energy profiles.


\clearpage
\section*{Acknowledgements}

B.~O. acknowledges useful discussions with Taylor Jacovich and Sarah Chastain, and thanks James Bauer and Quanzhi Ye for their assistance obtaining LDT observations. 
The authors would like to acknowledge the ATCA staff, in particular Mark Wieringa, for helpful discussions. 

This work was supported by the European Research Council through the Consolidator grant BHianca (grant agreement ID 101002761) and by the National Science Foundation (under award number 12850). The development of afterglow models used in this work was partially supported by the European Union Horizon 2020 Programme under the AHEAD2020 project (grant agreement number
871158). 

P.~B.'s research was supported by a grant (no. 2020747) from the United States-Israel Binational Science Foundation (BSF), Jerusalem, Israel. J.~G.’s research was supported by the Israel Science Foundation - National Natural Science Foundation of China joint research program under grant no. 3296/19. Research at Perimeter Institute is supported in part by the Government of Canada through the Department of Innovation, Science and Economic Development and by the Province of Ontario through the Ministry of Colleges and Universities. The material is based upon work supported by NASA under award number 80GSFC21M0002. 

Based on observations obtained with XMM-Newton, an ESA science mission with instruments and contributions directly funded by ESA Member States and NASA. This work made use of data from the NuSTAR mission, a project led by the California Institute of Technology, managed by the Jet Propulsion Laboratory, and funded by the National Aeronautics and Space Administration.  
Based on observations obtained with MASTER, supported by the Development Program of Lomonosov MSU and the UNU Astrophysical Complex of MSU-ISU (agreement EB-075-15-2021-675).
Based on observations obtained at the international Gemini Observatory, a program of NSF's OIR Lab, which is managed by the Association of Universities for Research in Astronomy (AURA) under a cooperative agreement with the National Science Foundation on behalf of the Gemini Observatory partnership: the National Science Foundation (United States), National Research Council (Canada), Agencia Nacional de Investigaci\'{o}n y Desarrollo (Chile), Ministerio de Ciencia, Tecnolog\'{i}a e Innovaci\'{o}n (Argentina), Minist\'{e}rio da Ci\^{e}ncia, Tecnologia, Inova\c{c}\~{o}es e Comunica\c{c}\~{o}es (Brazil), and Korea Astronomy and Space Science Institute (Republic of Korea). The Australia Telescope Compact Array is part of the Australia Telescope National Facility (grid.421683.a) which is funded by the Australian Government for operation as a National Facility managed by CSIRO. We acknowledge the Gomeroi people as the Traditional Owners of the Observatory site. This paper includes archived data obtained through the Australia Telescope Online Archive.

These results also made use of Lowell Observatory's Lowell Discovery Telescope (LDT), formerly the Discovery Channel Telescope. Lowell operates the LDT in partnership with Boston University, Northern Arizona University, the University of Maryland, and the University of Toledo. Partial support of the LDT was provided by Discovery Communications. LMI was built by Lowell Observatory using funds from the National Science Foundation (AST-1005313). 
This work made use of data supplied by the UK \textit{Swift} Science Data Centre at the University of Leicester. 
Some of the data used in this paper were acquired with the COATLI telescope and HUITZI imager at the Observatorio Astronómico Nacional on the Sierra de San Pedro Mártir. COATLI is funded by CONACyT (LN 232649, LN 260369, LN 271117, and 277901) and the Universidad Nacional Autónoma de México (CIC and DGAPA/PAPIIT IG100414, IT102715, AG100317, IN109418, IG100820, and IN105921) and is operated and maintained by the Observatorio Astronómico Nacional and the Instituto de Astronomía of the Universidad Nacional Autónoma de México. 
Based on observations collected at the Palomar Observatory with the 200-inch Hale Telescope, operated by the California institute of Technology, its divisions Caltech Optical Observations, the Jet Propulsion Laboratory (operated for NASA) and Cornell University. 
The \textit{HST} data used in this work was obtained from the Mikulski Archive for Space Telescopes (MAST). STScI is operated by the Association of Universities for Research in Astronomy, Inc., under NASA contract NAS5-26555. Support for MAST for non-\textit{HST} data is provided by the NASA Office of Space Science via grant NNX09AF08G and by other grants and contracts.

This research has made use of the XRT Data Analysis Software (XRTDAS) developed under the responsibility of the ASI Science Data Center (ASDC), Italy. 
This research has made use of the NuSTAR Data Analysis Software (NuSTARDAS) jointly developed by the ASI Space Science Data Center (SSDC, Italy) and the California Institute of Technology (Caltech, USA). 
This research has made use of data and/or software provided by the High Energy Astrophysics Science Archive Research Center (HEASARC), which is a service of the Astrophysics Science Division at NASA/GSFC. 

\newpage 

\begin{center}
{\LARGE \textbf{Supplementary Materials for A structured jet explains the extreme GRB 221009A}}
\end{center}

\begin{center}
B. O'Connor$^{1,2,3,4\ast}$, 
E. Troja$^{5,6\ast}$, 
G. Ryan$^{7}$, 
P. Beniamini$^{8,9}$, 
H. van Eerten$^{10}$, \\
J. Granot$^{8,9,1}$, 
S. Dichiara$^{11}$, 
R. Ricci$^{12,13}$, 
V. Lipunov$^{14}$, 
J. H. Gillanders$^{5}$, 
R. Gill$^{15}$, \\
M. Moss$^{1}$, 
S. Anand$^{16}$, 
I. Andreoni$^{17,3,4,\dagger}$, 
R. L. Becerra$^{18}$, 
D. A. H. Buckley$^{19,20}$, \\
N. R. Butler$^{21}$, 
S. B. Cenko$^{4,17}$, 
 A. Chasovnikov$^{14}$, 
J. Durbak$^{3,4}$, 
C. Francile$^{22,23}$, \\
E. Hammerstein$^{3}$, 
A. J. van der Horst$^{1}$, 
M. Kasliwal$^{16}$, 
C. Kouveliotou$^{1,2}$, \\
A. S. Kutyrev$^{3,4}$, 
W. H. Lee$^{24}$, 
G. Srinivasaragavan$^{3}$, 
V. Topolev$^{14}$, \\
A. M. Watson$^{24}$, 
Y.-H. Yang$^5$, 
K. Zhirkov$^{14}$
Corresponding author: oconnorb@gwmail.gwu.edu, eleonora.troja@uniroma2.it
\end{center}

\section*{Materials and Methods}

\noindent \textbf{Energetics and Rates}

GRB 221009A triggered \textit{Fermi}/GBM \cite{GCN32636} on 2022-10-09 at 13:16:59 UT. The GRB displayed an initial short pulse ($\sim$\,$40$ s) followed by a period of apparent quiescence, then a main emission episode consisting of two bright peaks,  at $T_0$\,$+$\,$225$~s and $T_0$\,$+$\,$260$~s, respectively.
A third, weaker peak is visible at $T_0$\,$+$\,$520$~s. 

\begin{figure*} 
\centering
\includegraphics[width=0.95\textwidth]{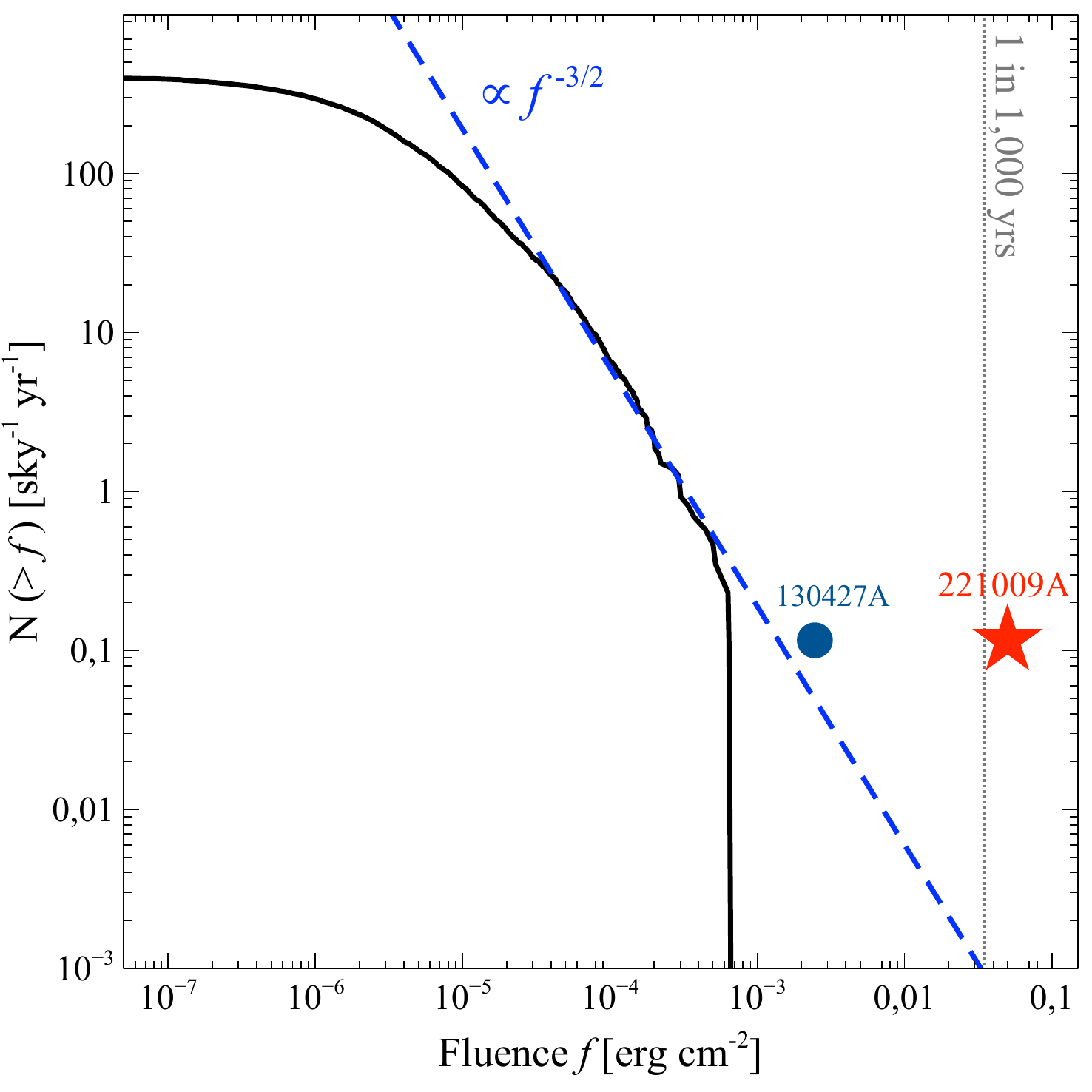}
\caption{\textbf{Fluence distribution of \textit{Fermi} GRBs.} 
We have normalized the number of bursts for the mission lifetime, its duty cycle, and field of view \cite{gbmcat}. 
At large fluences ($S\!\gtrsim$5$\times$10$^{-5}$ erg cm$^{-2}$) the distribution has a slope consistent with the euclidean one (–3/2), also shown for comparison (dashed line). 
The two most fluent GRBs are GRB 130427A (circle) and GRB~221009A (star). 
}
\label{fig:logNlogS}
\end{figure*}

Due to the GRB's immense brightness, the majority of satellites were saturated during the main emission episode.
This prevents us from carrying out standard analysis without careful corrections (Fermi GBM Team, in preparation). 
However, the burst fluence can be constrained using the \textit{Konus-Wind} spectrum at the onset of the main pulse ($T_0$\,$+$\,$180$\,$-$\,$200$ s).  This is described by a Band function \cite{Band1993} with $\alpha$\,$=$\,$-1.09$, $\beta$\,$=$\,$-2.6$ and peak energy $E_p$\,$\approx$\,$1$ MeV. 
By applying this model to the entire GRB prompt phase, a fluence of 
$\sim$\,$5.2\times 10^{-2}$ erg cm$^{-2}$ (20~keV\,$-$\,10 MeV) \cite{GCN32668} was derived between $0$\,$-$\,$700$ s after the initial trigger. 
As the spectrum of the brightest peaks is likely harder than the spectrum at the onset, this value places a conservative lower bound to the true GRB fluence,
and already makes GRB 221009A the brightest GRB ever detected by over an order of magnitude (Table \ref{tab:brightgrb}). 
Based on the fluence distribution of \textit{Fermi} bursts \cite{gbmcat}, the expected probability of observing a similar event is less than 1 in 1,000 years for a spatially homogeneous GRB population in Euclidean space (Fig. \ref{fig:logNlogS}).

At the redshift of $z$\,$=$\,$0.1505$ \cite{GCN32648xshooterz,GCN32686GTCz}, the observed fluence corresponds to an isotropic-equivalent energy of $E_{\gamma,\rm iso}$\,$\gtrsim$\,$3\times 10^{54}$ erg (1 keV\,$-$\,10 MeV), among a short list of the most energetic GRBs to date \cite{Atteia2017} and very similar to GRB 160625B \cite{Troja17}. 
Highly energetic ($\gtrsim$\,$10^{54}$) GRBs are intrinsically rare events. \textit{Swift} has detected approximately 11 of them during its entire lifetime \cite{Minaev2020}. 
As expected, most are in the redshift range $1$\,$\lesssim$\,$z$\,$\lesssim 3$, 
where the star formation (hence the GRB rate) peaks. 
Assuming a constant GRB formation rate over this redshift interval, we compute an all-sky volumetric rate of energetic GRBs: 
\begin{equation}
    R_\textrm{z} \approx \frac{N_{54}}{V} \frac{4\pi}{f_\textrm{z} \Omega_\textrm{BAT}}\frac{1}{\epsilon T} = 0.014^{+0.007}_{-0.006} \;\textrm{Gpc}^{-3} \textrm{yr}^{-1}
\end{equation}
where $N_{54}$\,$\approx$\,$8$ is the number of very energetic events detected within $1$\,$\lesssim$\,$z$\,$\lesssim$\,$3$ with Poissonian uncertainties derived from \cite{Gehrels86}, the available volume is $V$\,$=$\,$\int^3_1 \frac{dV}{dz}\frac{dz}{1+z}$\,$\approx$\,$300$ Gpc$^{-3}$ where $\frac{dV}{dz}$ is the comoving volume element \cite{Freedman2021},
$T$\,$\approx$\,$18$ yr is the current lifetime of the \textit{Swift} mission, $\epsilon\approx 78\%$ its duty cycle, and  $\Omega_{BAT}$\,$\approx$\,$2.2$ sr is the field of view of the \textit{Swift} Burst Alert Telescope (BAT) with partial coding $>$\,$10\%$. 
As these are very energetic events, we use an efficiency $f_\textrm{z}$\,$\approx$\,$0.8$ for obtaining a redshift measurement.

If we assume that the GRB rate scales as the star-formation rate \cite{HopkinsBeacom2006}, then the local rate of events is at least a factor of 10 lower, 
$R_\textrm{local}$\,$\approx$\,$0.001$ Gpc$^{-3}$ yr$^{-1}$, and the probability of seeing
a GRB as energetic as GRB 221009A within $z$\,$<$\,$0.15$ is only 1 in 1,000 yrs. 
Growing evidence shows that metallicity has a primary role in driving the GRB formation, and this could further lower the local rate by a factor $\approx$2 \cite{Palmerio2021,Ghirlanda2022}. 

A similar conclusion is reached by comparing GRB 221009A to the distribution of isotropic-equivalent GRB luminosities \cite{Salvaterra2012}. 
Utilizing the burst peak photon flux (10\,$-$\,1000 keV) from the preliminary GBM analysis, $2385$\,$\pm$\,$3$ ph s$^{-1}$ cm$^{-2}$ \cite{GCN32642}, and the \textit{Konus-Wind} spectral parameters reported above, we derive a peak gamma-ray luminosity of $L_{\gamma,\textrm{iso}}$\,$\approx$\,$9\times 10^{52}$ erg s$^{-1}$. 
In the simplest scenario, not including an evolution of the GRB rate with redshift or metallicity, we find that 
less than $\approx$\,0.1\% of GRBs have luminosities comparable to GRB 221009A, which translates in an all-sky rate of once in a decade across the visible Universe. For a local GRB  rate in the range  0.3\,$-$\,2 Gpc$^{-3}$ yr$^{-1}$,  we derive that the  rate of events as luminous as GRB 221009A is one 
every 300 to 1,100 years within $z$\,$\lesssim$\,$0.15$.

These independent analyses confirm that, due to its brightness and proximity to Earth, GRB~221009A is an exceptionally rare event.  

\noindent \textbf{Afterglow Temporal Evolution}

We model the afterglow lightcurves with a series of power-law segments, $F_{\nu}$\,$\propto$\,$t^{\alpha}$. 
In the X-ray band, our best fit model ($\chi^2$/dof$\,\approx$\,1.3 for 1679 dof) is a broken power-law with initial decay index $\alpha_{\textrm{X},1}$\,$=$\,$-1.52\pm0.01$, steepening to $\alpha_{\textrm{X},2}$\,$=$\,$-1.66\pm0.01$ after $t_{\textrm{break},X}$\,$=$\,$0.82\pm0.07$ d. A single power-law slope provides a significantly worse description of the data ($\chi^2$/dof\,$\approx$\,2.1). 

The OIR lightcurve, combining the $grizJHK$ and the MASTER \textit{C} filters, displays an initial power-law decay of $\alpha_{\textrm{OIR},1}$\,$=$\,$-0.88\pm0.05$, which steepens to $\alpha_{\textrm{OIR},2}$\,$=$\,$-1.42\pm0.11$ at around $t_{\textrm{break,OIR}}$\,$=$\,$0.63\pm0.13$ d ($\chi^2$/dof\,$\approx$\,1.3 for 95 dof). 
At times $t$\,$>$\,$10$ d, a slight deviation from this power-law is observed (Fig. \ref{fig: multiwavelength_lightcurve}), possibly caused by the contribution from the associated supernova.
By performing a joint-fit to the X-ray and OIR lightcurves, we derive a break time of $t_{\textrm{break,XOIR}}$\,$=$\,$0.79\pm0.04$ d ($\chi^2$/dof\,$\approx$\,1.3 for 1775 dof). This joint-fit results in a slightly steeper initial OIR decay of $\alpha_{\textrm{OIR},1}$\,$=$\,$-0.92\pm0.04$, whereas the other best-fit slopes remain consistent with our previous values.

The afterglow behaviour at radio energies is markedly different, suggesting that its evolution is decoupled from the higher energy data. 
Our dataset starts 6 d after the GRB trigger and 
follows the afterglow evolution up to 110 d in multiple frequencies. 
During the time interval 6-41 d, 
we derive a power-law decay with slope $-0.76\pm0.08$ 
at 16.7 GHz (Fig. \ref{fig: multiwavelength_lightcurve}),
much shallower than the simultaneous OIR and X-ray light curves. 
A consistent behavior is observed at 21.2 GHz, whereas higher ($>$30 GHz) frequencies show a possible chromatic steepening after $\approx$30 d. 
A comparison with the lower frequency data (5.5 GHz) 
shows that the simple power-law model overpredicts the radio flux at $t\!\lesssim$\!1 d, requiring the presence of a temporal break at early times.

\noindent \textbf{Afterglow Spectral Properties}

\noindent \textbf{\textit{X-rays}}

We model the afterglow spectra in each energy band (X-ray, OIR, radio) using a simple power-law function, $F_{\nu}$\,$\propto$\,$\nu^{\beta}$.
The best fit was found by minimizing the Cash statistic \cite{Cash1979} within \texttt{XSPEC v12.12.0} \cite{Arnaud1996}. 
The effects of absorption were included using the \texttt{XSPEC} model
\texttt{tbabs$*$ztbabs$*$pow} with fixed redshift $z$\,$=$\,$0.1505$ \cite{GCN32648xshooterz}, ISM abundance table from \cite{Wilms2000}, and photoelectric absorption cross-sections from \cite{Verner1996}.
The GRB sightline intercepts dense clouds along the Galactic plane, as shown
by the bright dust scattering echoes at X-ray energies
\cite{GCN32680, Williams2023}. We probed the absorbing column in the GRB direction by extracting multiple spectra of the ring located 6 arcmin from the GRB position at $T_0$\,$+$\,$1.2$ d (ObsID: 01126853005). 
They are well described by a power-law with photon index 
$\Gamma$\,$=$\,$4.3^{+0.6}_{-0.5}$ and an absorbing column of $N_H$\,$=$\,$(2.1^{+0.2}_{-0.6})\times10^{22}$ cm$^{-2}$. 
We used the latter value as our estimate of the 
Galactic hydrogen column density. 

The soft X-ray spectra (0.3-10 keV) display an initial hard spectral index of $\beta_\textrm{X}$\,$=$\,$-0.65\pm0.02$ at 1 hr, which is seen to soften with time to  $\beta_\textrm{X}$\,$=$\,$-0.85\pm0.03$ at 5 hr, 
and $\beta_\textrm{X}$\,$=$\,$-0.92\pm0.01$ at 32 d.
A similar trend is measured in the \textit{NuSTAR} data ($3-79$ keV), which also display a spectral softening between $-0.81\pm0.01$ at 1.8 d and $-1.10\pm0.17$ at 32 d. The latter value is consistent with the initial spectral index, $\beta_\textrm{BAT}$\,$=$\,$-1.08\pm0.03$, determined by \textit{Swift} BAT \cite{Krimm2022,Williams2023}.

\noindent \textbf{\textit{Optical and Infrared}}

\begin{figure} 
\centering
\includegraphics[width=0.8\columnwidth]{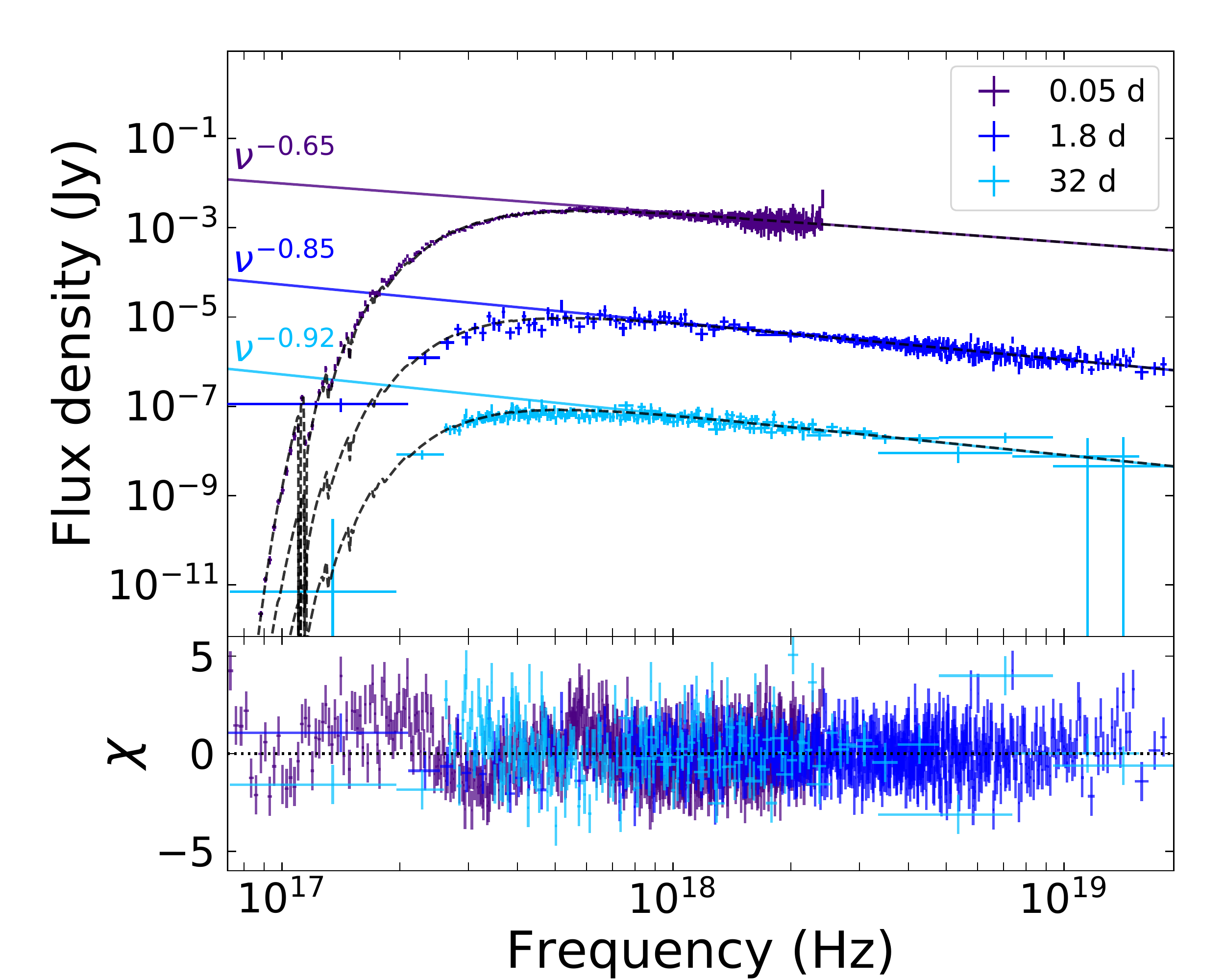}
\includegraphics[width=0.8\columnwidth]{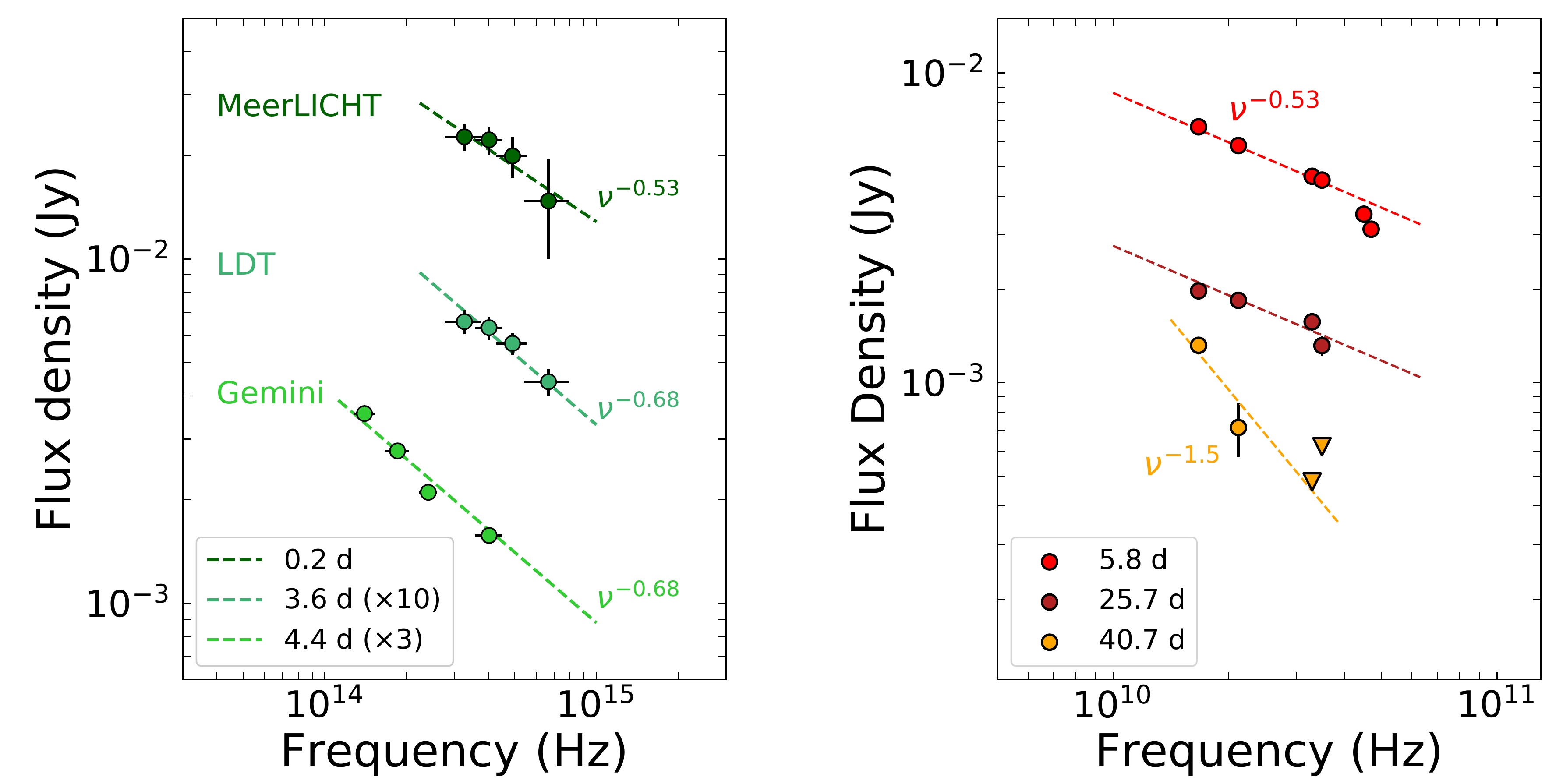}
\caption{\textbf{Multi-epoch spectra and SEDs of GRB 221009A.} 
Panel A (top): X-ray spectra of GRB 221009A fit an absorbed power-law model: \textit{Swift} at 0.05 d, \textit{Swift} and \textit{NuSTAR} at 1.8 d, \textit{XMM-Newton} and \textit{NuSTAR} at 32 d. 
Panel B (bottom left): Spectral energy distributions of the optical and infrared data fit with a simple power-law model. The OIR data have been corrected for Galactic extinction $E(B-V)$\,$=$\,$1.32$ mag \cite{Schlafly2011}.
Panel C (bottom right): Spectral energy distributions of the radio data ($16.7$\,$-$\,$47$ GHz).  
}
\label{fig: X_SEDs}
\end{figure}

The OIR data were modeled within \texttt{XSPEC} 
using the model \texttt{redden$*$zdust$*$pow} with fixed redshift $z$\,$=$\,$0.1505$. 
Using a Galactic extinction of $E(B-V)$\,$=$\,$1.32$ mag \cite{Schlafly2011}, 
we derive a negligible intrinsic extinction $E(B-V)_z$\,$<$\,$0.1$ mag at the 3$\sigma$ confidence level, 
a spectral index $\beta_\textrm{OIR}$\,$=$\,$-0.53\pm0.10$
between $0.2$\,$-$\,$0.5$ d, and a steeper index $\beta_\textrm{OIR}$\,$=$\,$-0.68\pm0.05$ after 1.7 d 
(see Fig. \ref{fig: X_SEDs}). 
These values are consistent with the 
spectral index of the early ($\sim$\,1 hr) X-ray afterglow.

\noindent \textbf{\textit{Radio}}

Our dataset spans the frequency range between 5.5 GHz and 47 GHz. Our second epoch (14.7 d post-burst) was not included in the spectral analysis due to the large systematic uncertainty at higher frequencies. As shown in Fig. \ref{fig: X_SEDs}, the spectra at 5.8 and 25.7 d can be described by a power-law with spectral index $\beta_\textrm{R}=0.53\pm0.17$. 
There is possible evidence for a turn-over of the radio spectrum at 40.7 d, suggesting that the component powering the low-frequency radio emission is not contributing to the optical and X-ray flux.

\noindent \textbf{Theoretical Modeling}

\noindent \textbf{\textit{Standard Jet Model}}

We began by exploring the simple scenario of a relativistic fireball \cite{Meszaros1997,Sari1998,Wijers1999,Granot2002} propagating into an ambient medium with density of the form $\rho_{\textrm{ext}}(r)$\,$=$\,$A\, r^{-k}$.
The shock-accelerated electrons have an energy distribution of the form $N(E)$\,$\propto$\,$E^{-p}$, and
cool via synchrotron radiation with a broadband spectrum described by three break frequencies: the cooling frequency $\nu_c$, the characteristic frequency $\nu_m$, and the self-absorption frequency $\nu_a$. We focus mainly on the properties of the X-ray and OIR afterglow, as our analysis shows that the radio emission is dominated by a different component. 

Our observational constraints include (\textit{i}) the significant spectral evolution of the optical and X-ray emission during the first 24 hours since the GRB, and (\textit{ii}) the nearly achromatic steepening of the optical and X-ray lightcurves at around 0.8 d. 
A simple explanation for the spectral and temporal evolution of the X-ray counterpart is the passage of the cooling break. The observed change in X-ray spectral index, $\Delta\beta_{\rm X}$\,$=$\,$0.46\pm0.07$ between 0.05 and 40 d, points to a decreasing cooling break. The synchrotron cooling frequency is expected to change with time as $\nu_c\propto t^\frac{3k-4}{2k-8}$. This constraint implies $k$\,$<$\,$4/3$ and rules out a wind-like environment with density profile $k$\,$=$\,$2$ or steeper. However, if $\nu_c$\,$\lesssim$\,$\nu_{\rm X}$ at 1 d, the X-ray temporal slope after the break, $\alpha_{\rm X}$\,$=$\,$-1.66$, would require a steep $p$\,$=$\,$(2-4 \alpha_{\rm X})/3$\,$\approx$\,$2.88$, hence a soft spectral index $\beta_{\rm X}$\,$=$\,$p/2$\,$\approx$\,1.4 inconsistent with the observed spectral shape by $\approx$\,$5\sigma$. 

A possible solution could be to include a time-dependent evolution of the shock microphysical parameters \cite{Filgas2011,Troja12,Maselli13}: 
$\varepsilon_e$ is the fraction of the burst kinetic energy $E_\textrm{K}$ in electrons and $\varepsilon_B$ is the fraction in magnetic fields. 
To explain the slope of the X-ray lightcurve after the break, 
we require that the time-dependence of the microphysical parameters adds an additional $t^{-0.51}$ to the temporal decay above $\nu_c$ (assuming $p$\,$=$\,$2.2$). 
As the flux above $\nu_c$ has a very shallow dependence on $\varepsilon_B$ ($\propto$\,$\varepsilon_B^{1/20}$ for $p$\,$=$\,$2.2$), it is more practical to consider a time-evolution of $\varepsilon_e$. 
We find that $\varepsilon_e$\,$\propto$\,$t^{-0.425}$ 
would reproduce the X-ray temporal slope, but also 
drive a fast evolution of the spectral peak toward lower frequencies, $\nu_m$\,$\propto$\,$t^{-2.35}$, severely overpredicting the observed radio flux. 
Moreover, it does not account for the nearly simultaneous steepening of the optical emission. 
We therefore conclude that the passage of the cooling break across the X-ray band can explain some of the observed properties (e.g. the spectral softening) but does not entirely account for the steep temporal decay.

A straightforward explanation for the nearly achromatic temporal break at 0.8 d is a geometrical effect \cite{Rhoads1999,Sari1999}. 
The beamed geometry of the outflow causes the afterglow to decay at a faster rate once the jet-edges become visible, with a change in temporal slope of $\Delta\alpha$\,$=$\,$(3-k)/(4-k)$ in absence of lateral spreading, for a uniform sharp-edged (or `top-hat') jet. To be consistent with the X-ray measurements of $\Delta\alpha_{\rm X}$\,$=$\,$0.14\pm0.02$, this model requires a steep density profile $k$\,$\approx$\,$2.8$ in disagreement with the limit $k$\,$<$\,$4/3$ derived above. 

\begin{figure} 
\centering
\includegraphics[width=0.95\columnwidth]{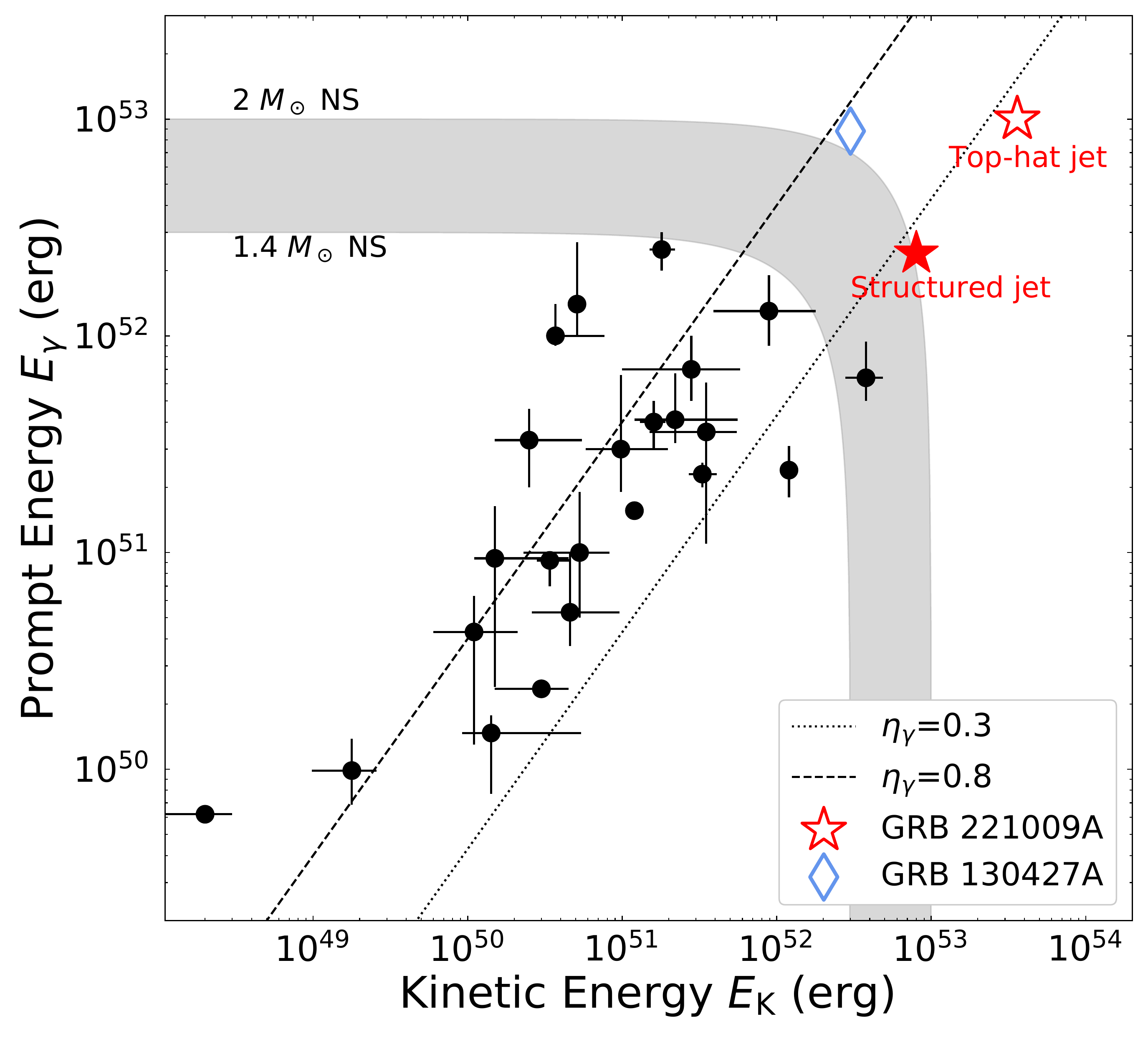}
\caption{\textbf{Collimation corrected kinetic energy $E_\textrm{K}$ versus prompt gamma-ray energy $E_\gamma$.} We have displayed a sample of long GRBs from the literature \cite{Cenko2011,DePasquale2016}. The empty red star displays the lower limit to the energy of GRB 221009A in the top-hat jet scenario and the filled red star in the structured jet case. The shaded gray regions shows a range of allowed values for magnetar central engines \cite{USOV1992,Thompson04}. The black lines show a gamma-ray efficiency of $\eta_\gamma$\,$=$\,$30$\,$-$\,$80\%$.  
}
\label{fig: Beamed_energy}
\end{figure}

If the temporal break at 0.8 d is not due to the collimation of the GRB outflow, the required energetics to power the burst are challenging to reproduce. 
The X-ray lightcurve evolves as a power-law up to at least 80 days after the trigger (at the time of writing). 
Therefore, we can put an observational lower limit on $t_{\rm j}$\,$>$\,$80$ d, which leads to a lower limit on both the jet opening angle $\theta_{\rm j}$ and the collimation-corrected energy of the GRB. 
Assuming $k=0$ (uniform medium) and a redshift $z=0.1505$, 
we derive: 
\begin{equation}
    E_\textrm{K}>4\!\times \!10^{53}\mbox{ erg}\left(\frac{t_{\rm j}}{80\mbox{ d}}\right)^{3/4}\left(\frac{1+z}{1.15}\right)^{-3/4} \left(\frac{E_{\rm K,iso}}{10^{55}\; \textrm{erg}}\right)^{3/4}\left(\frac{n}{1\; \textrm{cm}^{-3}}\right)^{1/4}
\end{equation}
where 
$n$ is the circumburst density for a uniform medium. 
Such a high beaming corrected energy 
would be an outlier in the GRB energy distribution (Fig. \ref{fig: Beamed_energy}; \cite{Cenko2011}).

\noindent \textbf{\textit{A Structured Jet}}

The standard assumption that GRB jets have a constant energy $dE_\textrm{K}/d\Omega$ within the core of the jet is likely an oversimplification, and a structured jet naturally arises as the GRB breaks out of its stellar environment \cite{Gottlieb2021,Gottlieb2022}. Here, we consider a GRB jet with a broken power-law structure (Fig. \ref{fig: schematic}) defined by:
\begin{eqnarray}
\label{energydist}
    \frac{dE_\textrm{K}}{d\Omega}\propto\left\{ 
\begin{array}{ll}\theta^{-a_1}& \textrm{for}\;\theta<\theta_{b}, \\ \\
\theta^{-a_2}  & \;\textrm{for}\;\theta_b<\theta<\theta_{s},
\\ \\
\theta^{-a_3}  & \;\textrm{for}\;\theta>\theta_{s},
\end{array} \right.  
\end{eqnarray}
where 
$a_1$\,$<$\,$a_2$\,$<$\,2, $a_3$\,$\rightarrow$\,$\infty$, and an observer angle $\theta_{\rm obs}$\,$<$\,$\theta_b$. This jet is therefore comprised of an initial shallow slope followed by a steeper lateral profile that becomes visible when $\Gamma\approx \theta_{b}^{-1}$, where $\theta_{b}$ is the width of the initial shallow profile. In this case, the flux and frequency evolution is dictated by the lateral structure, leading to a shallow angular structure dominated emission (sASDE) phase (see Tables 1 and 2 of \cite{BGG2022}). 
The flux evolution is initially the same as for a spherical outflow with an isotropic-equivalent energy corresponding to the value along the line of sight. This lasts up to a time, $t_{\rm sph}$, when the Lorentz factor along the line of sight has decelerated to $\theta_{\rm obs}^{-1}$. At this point, the lateral structure becomes visible and the sASDE phase begins. In this regime, the flux in an ISM environment is determined by:
\begin{eqnarray}
\label{Fluxdecay}
    F_{\nu}\propto\left\{ \begin{array}{ll}t^\frac{3(a+2p-2)}{(a-8)}& \:\textrm{for}\:\nu_m<\nu<\nu_c, \\ \\
t^\frac{2(a+3p-2)}{a-8}  & \:\textrm{for}\:\nu_c<\nu. 
\end{array} \right.
\end{eqnarray}
This phase lasts until $t_{\rm b}$\,$\approx$\,$0.8$ d, which is the time at which $\Gamma$\,$=$\,$\theta_{b}^{-1}$. For $t$\,$>$\,$t_{\rm b}$, the flux evolves in a similar way (i.e., governed by Equation \ref{Fluxdecay}), but with $a_1$\,$\rightarrow$\,$a_2$. For angles $\theta$\,$>$\,$\theta_s$, a steep jet-break is expected at late-times ($t_\textrm{j}$\,$>$\,$80$ d) if the jet is still relativistic \cite{Frail2000, VanEerten2012}. 

We consider a scenario where $t_{\rm sph}$\,$<$\,$1$ hr, such that the temporal decay is provided by Equation \ref{Fluxdecay} for the entire time period of our observations ($0.05$\,$-$\,$80$ d). We find a solution for $p$\,$=$\,$2.2$, $k$\,$=$\,$0$, $a_1$\,$=$\,$0.75$, and $a_2$\,$=$\,$1.15$, which yields an initial pre-break slope of 
$\alpha_{\textrm{X},1}$\,$=$\,$-1.55$ that steepens to $\alpha_{\textrm{X},2}$\,$=$\,$-1.67$ after $\sim$\,0.8 d, assuming that $\nu_c$\,$<$\,$\nu_{\rm X}$. The OIR slope after the break is $\alpha_{\textrm{OIR},2}$\,$=$\,$-1.47$ for $\nu_m$\,$<$\,$\nu_\textrm{OIR}$\,$<$\,$\nu_c$, while the early OIR data ($<$\,$1$ d) is dominated by emission from a different component of emission, likely a reverse shock.  

Motivated by this solution, we modeled the afterglow spectral energy distributions (SEDs) from radio to X-ray wavelengths with a phenomenological model combining two components, a forward shock (FS) and reverse shock (RS). The FS closure relations are governed by the jet structure (Equation~\ref{Fluxdecay}).

The RS evolution 
in a structured outflow has not been sufficiently developed yet for inclusion  in our study. 
We therefore adopt standard prescriptions
of a thin-shell model parameterized by the power-law index $g$ of the Lorentz factor distribution with radius $\Gamma\propto R^{-g}$ \cite{Kobayashi2000}. 
Each component was allowed to have an independent electron spectral slope $p$, and was parameterized by the locations of $\nu_a$, $\nu_m$, and $\nu_c$ at a reference time of 1 d as well as the peak flux $F_{\nu,\mathrm{max}}$ at that time. 
The best fit was obtained by simple $\chi^2$ minimization on the extinction-corrected afterglow data (Fig. \ref{fig: multiwavelength_SED}). 

We find that the FS reproduces the full X-ray lightcurve and the OIR data after 0.8 d (Fig. \ref{fig: multiwavelength_SED}). In this model, the cooling break $\nu_{c,\textrm{FS}}$ moves through the X-ray band, producing the observed spectral softening.  
The injection frequency $\nu_{m,\textrm{FS}}$ is constrained to be close to the soft X-ray band at early times, as otherwise the FS severely overpredicts the radio emission at later times ($>$\,$14$ d). 
A jet model with a single slope for the power-law energy distribution $dE_\textrm{K}/d\Omega$\,$\propto$\,$\theta^{-a}$ with $a$\,$\approx$\,$1$ can reproduce these observations. 
However, a broken power-law energy structure is well motivated by simulations \cite{Gottlieb2021} of GRB jets, which display an evolution of the jet's angular energy distribution from shallower to steeper slopes. 
For a jet with structure given by $a_1$\,$\approx$\,$0.75$ and $a_2$\,$\approx$\,$1.15$, 
we derive $\nu_{c,\textrm{FS}}$\,$\approx$\,$1.7\times10^{18}$ Hz, $\nu_{m,\textrm{FS}}$\,$\approx$\,$3.6\times10^{14}$ Hz, 
$F_{\nu,\textrm{max,FS}}$\,$\approx$\,$5.9$ mJy, and $p$\,$\approx$\,$2.25$. The self-absorption frequency $\nu_{a,\textrm{FS}}$\,$\lesssim$\,$10^{10}$ Hz 
remains unconstrained. 

Afterglow modelling depends on a large number of physical parameters, larger than the number of observed constraints. Many values of the physical parameters, varying over orders of magnitude, will produce nearly identical afterglow emission.  Using fiducial values of $E_\textrm{K,iso}$\,$=$\,$10^{55}$ erg, $n$\,$=$\,$1$ cm$^{-3}$, and $\theta_\textrm{obs}$\,$=$\,$0.01$ rad, and inverting the values of $F_{\nu,\textrm{max,FS}}$, $\nu_{m,\textrm{FS}}$, $\nu_{c,\textrm{FS}}$, we find a solution for $\varepsilon_e$\,$=$\,$0.17$, $\varepsilon_B$\,$=$\,$4.4\times10^{-6}$, and an electron participation fraction $\xi_N$\,$=$\,$0.015$.

\begin{figure*}
\centering
\includegraphics[width=0.95\textwidth]{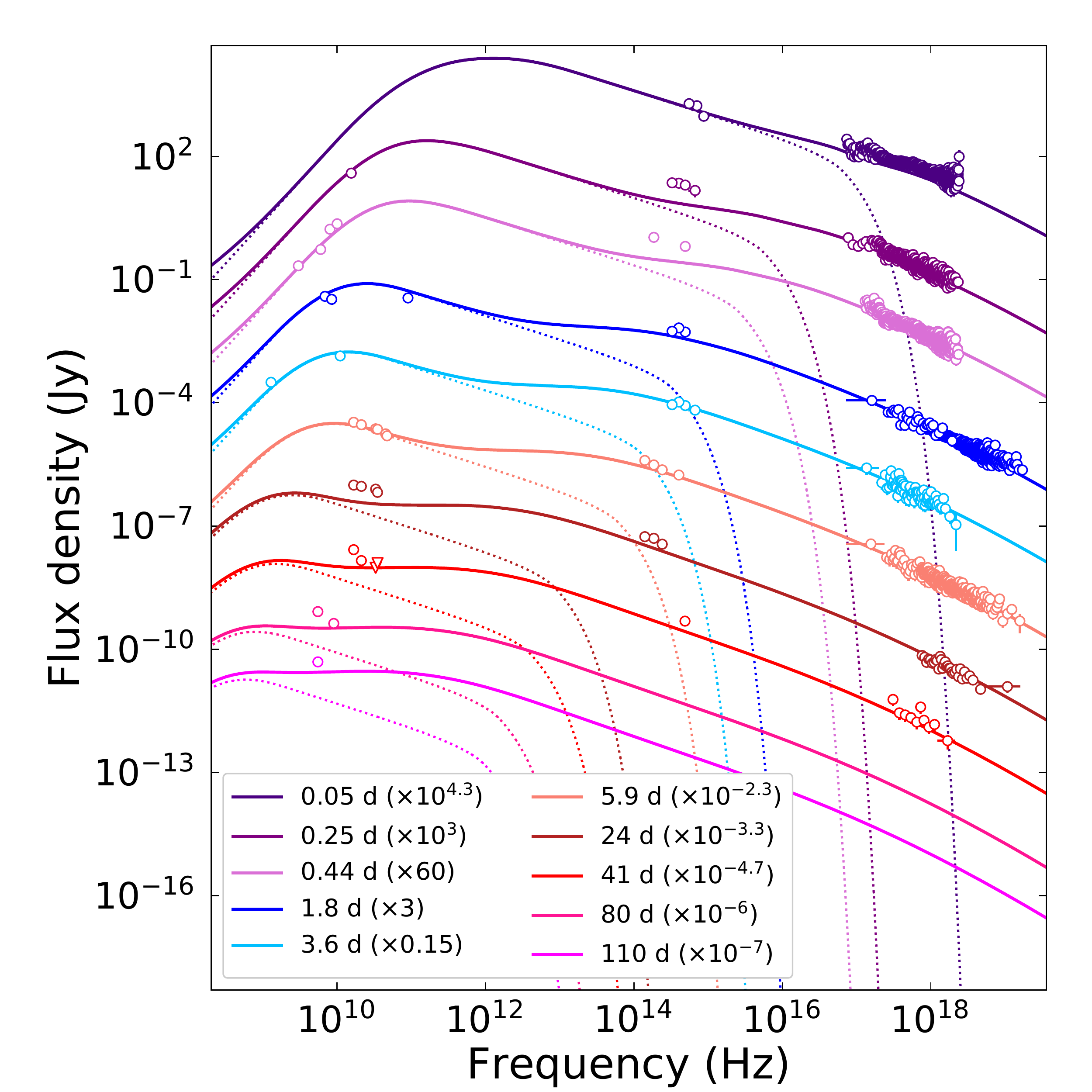}
\caption{\textbf{Multi-epoch broad-band SEDs of GRB 221009A.} We have modeled the data by the combination (solid line) of a forward and reverse shock (dotted line). A single reverse shock evolved following the thin-shell closure relations \cite{Kobayashi2000} cannot reproduce the early optical and late radio emission. 
The data is corrected for extinction and absorption. 
}
\label{fig: multiwavelength_SED_2}
\end{figure*}

As expected, we find that the standard thin-shell RS is not capable of reproducing the phenomenology of a structured outflow. Even for large values of the parameter $g$\,$\gtrsim$\,$3$, 
this model cannot account for the early optical emission and, at the same time, reproduce the shallow decay of the radio afterglow (Fig. \ref{fig: multiwavelength_SED_2}).
In order to capture this behavior, we require at least two separate RS components (Fig. \ref{fig: multiwavelength_SED}), with the first RS dominating at $<$\,$10$ d capable of explaining the early radio and OIR data. Alternatively, ad-hoc solutions such as energy injection and/or variability of the shock microphysics may be introduced to slow down the RS evolution. 

Similar challenges were encountered in the modeling of other bright GRBs \cite{Kangas21}. 
For example, a two component model was explored in the case of GRB 030329A \cite{Berger2003}, GRB 130427A \cite{van_der_Horst2014} and GRB 190829A \cite{Sato21,Dichiara2022}
whereas time-evolving microphysics were favored in the case of GRB 190114C \cite{Misra2021}. 
GRB 221009A adds to the growing sample of bursts deviating from the basic RS scenario, and motivates further extension of the standard model by incorporating a broader set of jet angular structures.



\noindent \textbf{Implications of a Structured Jet}

\noindent \textbf{\textit{Energetics}}

An advantage of the shallow structured jet scenario suggested above is that it reduces the energy requirements compared to steep jet models. 
The viewing angle to the burst $\theta_{\rm obs}$ is directly related to the earliest time at which the afterglow begins evolving according to the inner (shallow) slope of the energy profile, $t_{\rm sph}$ (i.e., $E_{\rm K,iso}$\,$\propto$\,$\theta^{-a_1}$, see Equation \ref{Fluxdecay}) 
and can be expressed as:
\begin{equation}
\label{thetaobs}
    \theta_{\rm obs}=\left(\frac{(3-k)E_{\rm K,iso}(\theta_{\rm obs})}{4\pi A \,2^{3-k} c^{5-k}}\right)^{\frac{1}{2k-8}}t_{\rm sph}^{\frac{3-k}{8-2k}} 
\end{equation}
which, for $k$\,$=$\,$0$ (as favoured by the spectral evolution), implies $\theta_{\rm obs}$\,$\lesssim$\,$0.016$ rad for $t_{\rm sph}$\,$<$\,$0.05$ d. 
For a shallow structured jet, the maximum polar angle of the jet viewable to an observer at $\theta_{\rm obs}$ evolves in time according to:
\begin{eqnarray}
    \theta(t)\!=\!\theta_{\rm obs}\left\{ \begin{array}{ll}(t/t_{\rm sph})^{\frac{3-k}{8-2k-a_1}} & \:\textrm{for}\:t_{\rm sph}<t<t_{\rm b} ,\\ \\(t_{\rm b}/t_{\rm sph})^{\frac{3-k}{8-2k-a_1}}(t/t_{\rm b})^{\frac{3-k}{8-2k-a_2}}  \quad  & \:\textrm{for}\:t>t_{\rm b},
\end{array} \right.
\end{eqnarray}
where we adopt $a_1$\,$=$\,$0.75$ and $a_2$\,$=$\,$1.15$. 
This is related to the angle $\theta_{\rm b}$ below which $E_{\rm K,iso}$ evolves as $\theta^{-a_1}$ and above which as $\theta^{-a_2}$. 
Therefore, $\theta_{\rm b}$ can be expressed as:
\begin{eqnarray}
  & \theta_{\rm b} =\left(\frac{(3-k)E_{\rm K,iso}(\theta_{\rm obs})}{4\pi A\, 2^{3-k} c^{5-k}}\right)^{\frac{1}{2k-8+a_1}}t_{\rm b}^{\frac{3-k}{8-2k-a_1}}\theta_{\rm obs}^{\frac{-a_1}{8-2k-a_1}} \nonumber \\ 
  & = 0.057\;{\rm rad} \left(\frac{E_{\rm K,iso}}{10^{55}\,\textrm{ erg}} \right)^{-0.14}   \left(\frac{n}{1\,\textrm{ cm}^{-3}}\right)^{0.14}   \left(\frac{\theta_{\rm obs}}{0.01}\right)^{-0.10} \left(\frac{t_{\rm b}}{0.8\, \textrm{ d}}\right)^{0.4} \left(\frac{1+z}{1.15}\right)^{-0.4}.
\end{eqnarray}
As we do not observe a steep break (i.e. traditional jet-break with $F_{\nu}$\,$\propto$\,$t^{-p}$) in the lightcurve out to $t_{\rm j}$\,$>$\,$80$ d we can set a lower limit to the opening angle $\theta_s$ (see Equation \ref{energydist}) out to which the relation $E_{\rm K,iso}$\,$\propto$\,$\theta^{-a_2}$ extends. This in turn allows us to derive a lower limit to the collimation-corrected kinetic energy in the jet $E_{\rm K}$. Using $\theta_{\rm obs}$ and $\theta_{\rm b}$, these relations are provided by: 
\begin{eqnarray}
  &  \theta_{\rm s} > \theta(t)=\theta_{\rm b} \left(\frac{t_{\rm j}}{t_{\rm b}}\right)^{\frac{3-k}{8-2k-a_2}}\\
 & E_{\rm K} = \frac{a_3-a_2}{(2-a_2)(a_3-2)}\theta(t)^2 E_{\rm K, iso}(\theta_{\rm obs}) \left(\frac{\theta_{\rm b}}{\theta_{\rm obs}}\right)^{-a_1}\left(\frac{\theta(t)}{\theta_{\rm b}}\right)^{-a_2}, 
\end{eqnarray} 
which for $\theta(t)$\,$=$\,$\theta_s$, $k$\,$=$\,$0$,  $a_1$\,$=$\,$0.75$, $a_2$\,$=$\,$1.15$, and $a_3$\,$\rightarrow$\,$\infty$, becomes:
\begin{eqnarray}
\label{thetas}
& \theta_{\rm s} > 0.4\;{\rm rad} \left(\frac{E_{\rm K,iso}}{10^{55}\,\textrm{erg}} \right)^{-0.14}   \left(\frac{n}{1\,\textrm{ cm}^{-3}}\right)^{0.14}    \left(\frac{\theta_{\rm obs}}{0.01}\right)^{-0.10} \left(\frac{t_{\rm b}}{0.8\, {\rm  d}}\right)^{-0.04} \left(\frac{1+z}{1.15}\right)^{0.04} \left(\frac{t_{\rm j}}{80\, {\rm  d}}\right)^{0.44}\\
& E_{\rm K} = 8\times10^{52} \left(\frac{E_{\rm K,iso}}{10^{55}\,\textrm{erg}}\right)^{0.83}          \left(\frac{n}{1\,\textrm{ cm}^{-3}}\right)^{0.17} \!\left(\frac{\theta_{\rm obs}}{0.01}\right)^{0.62}\!  \left(\frac{t_{\rm b}}{0.8\,{\rm d}}\right)^{0.14}\! \left(\frac{t_{\rm j}}{80\,{\rm d}}\right)^{0.37} \left(\frac{1+z}{1.15}\right)^{-0.51}  
\end{eqnarray}
Formally, this limit decreases for smaller viewing angles $\theta_\textrm{obs}$. However, considering that $t_{\rm sph}$ has to be greater than the prompt duration of the GRB, $\theta_{\rm obs}$ can not decrease by much.  
The required energy is reduced compared to the standard jet case ($4\times10^{53}$ erg). Furthermore, the energy has a significantly shallower dependence on the time of the steep jet-break $t_{\rm j}^{0.37}$ compared to $t_{\rm j}^{3/4}$ for a top hat jet. In other words, as the length of time over which we do not observe a steep jet-break increases, the more energetically favorable the structured jet model becomes. 

A structured jet with extended wings will take even longer to establish causal contact across the jet surface than a top-hat jet, which can help to explain the lack of post-jet break dynamics observed for GRB 221009A and similar events. In the central engine frame, a relativistic sound wave travelling along the jet surface between edge and tip will move along with velocity $\beta_\theta$\,$=$\,$1 / (2 \Gamma)$ for a jet with local shock Lorentz factor $\Gamma$ \cite{VanEerten2013}. If $\Gamma$\,$\propto$\,$\theta^{-a_1/2}$, causal contact out to angle $\theta$ will occur once $\Gamma(\theta)$\,$=$\,$(1-a_1/2)/(3 \theta)$\,$\lesssim$\,$1$ (for example, $\theta$\,$=$\,$0.4$ rad, $a_1$\,$=$\,$0.75$, 
$\Gamma$\,$=$\,$0.52$\,$<$\,$1$ shows a clear breakdown of the assumption of relativistic dynamics; a steepening to a jet structure slope $a_2$ at intermediate angle would lead to an even lower $\Gamma$). The non-relativistic lightcurve slope above the cooling break (e.g. in the X-rays), is given by $t^{(4-3p)/2}$ \cite{Frail2000, VanEerten2012}, which can be very similar to the relativistic slope of a shallow structured jet. 

\noindent \textbf{\textit{Rate of Events}}

We have demonstrated that GRB 221009A stands out compared to other long-duration GRBs in terms of both its energetics and close proximity. The shallow flux decay in the X-ray and OIR until very late times is interpreted as evidence for a shallow jet structure. We therefore suggest that GRB 221009A, and other nearby GRBs without steep jet breaks (GRB 130427A, GRB 180720B, GRB 190114C, and GRB 190829A), imply the existence of a sub-class of energetic GRBs with shallow jet structures. 

This sub-class, due to their shallow angular profiles, has a different effective beaming compared to the typical GRB population, which affects their observed rate. 
On the one hand, the large inferred value of $\theta_{\rm s}$\,$\approx$\,$0.4$ rad derived for GRB 221009A in Equation \ref{thetas} is significantly larger than typical opening angles derived for long GRBs, i.e., $\theta_{\rm j}$\,$\approx 0.1$ rad \cite{Goldstein2016,Wang2018}, where jet breaks are easier to observe for energetic and narrow jets. This would suggest, that even a relatively small intrinsic rate associated with the sub-population of shallow jets might be over-represented in the observed data, approximately by a factor of $(\theta_{\rm s}/\theta_{\rm j})^2\approx16$. However, the low derived value of the viewing angle to GRB 221009A, $\theta_{\rm obs}$\,$\lesssim$\,$0.016$ rad (Equation \ref{thetaobs}) is in tension with this suggestion. In particular, due to the larger solid angles associated with greater viewing angles, for each burst like GRB 221009A that is viewed from  $\lesssim$\,$\theta_{\rm obs}$, there should be (on average) $\sim$\,$625$ GRBs viewed from $\lesssim$\,$\theta_{\rm s}$. If we take a shallow angular profile of the kinetic energy, $E_{\rm K,iso}$\,$\propto$\,$\theta^{-a}$ with $a$\,$\approx$\,$0.9$ 
between $\theta_{\rm obs}$ and $\theta_{\rm s}$, then bursts viewed from $\sim$\,$\theta_{\rm s}$ might be expected to have a gamma-ray fluence that is roughly $20$ times smaller than that of GRB 221009A. In other words if the observed rate of GRB 221009A is about 1 in 1000 
years, then roughly once in 1.5 yr 
we should be detecting bursts which are $\sim$\,20 times less fluent. As shown in Fig. \ref{fig: bright_xrt_lightcurve} and Fig. \ref{fig:logNlogS}, this is clearly in contradiction with Fermi and BATSE observations. This suggests that, even if GRB 221009A-like jets have shallow profiles extending up to large latitudes, their gamma-ray production might be restricted to a much narrower range (up to some $\theta_{\gamma}$\,$\ll$\,$\theta_{\rm s}$). Such a possibility is expected if there is even a relatively small reduction in the bulk Lorentz factor of the outflows with $\theta$ \cite{Granot2018,BN2019}. \cite{BN2019} have argued that this is a limiting factor in the detectability of long GRBs based on various observational lines of evidence. To conclude, the intrinsic rate of GRB 221009A-like jets is strongly dependent on the effective opening angle for gamma-ray production, and with only one well constrained event of this type the intrinsic rate of such bursts remains largely unconstrained. Nonetheless, if indeed $\theta_{\gamma}$\,$\ll$\,$\theta_{\rm s}$ as suggested by the discussion above, then there should be a large population of similar jets which would have produced little or no gamma-rays, despite having been viewed from $\theta$\,$<$\,$\theta_{\rm s}$ and therefore corresponded to very bright and initially fast evolving afterglows in all wavelengths. The existence of these on-axis orphan afterglows of shallow jets can be constrained using transient surveys \cite{NP2003}. 

Assuming that the prompt GRB emission is produced in an optically thin region of the outflow, the dissipation radius inferred from the prompt emission variability timescale $\delta t$ can be compared to the photospheric radius in order to place a constraint on the minimum outflow Lorentz factor placing emission beyond the photospheric radius. Following \cite{LambKobayashi2016}, this requirement translates to $\Gamma$\,$\gtrsim$\,$505 \left( E / 10^{55} \textrm{ erg} \right)^{1/5} \left(\delta t / 0.1 \textrm{ s} \right)^{-2/5}$ 
for a jet of isotropic-equivalent energy $E$. 
The atypical jet structure inferred for GRB 221009A, with a very narrow core (a tip) embedded within shallow power-law profile, implies a far smaller jet surface area detectable in prompt emission. Assuming an inner jet Lorentz factor profile $\Gamma$\,$=$\,$\Gamma_\textrm{tip} \left( \theta / \theta_{\rm j} \right)^{-a_1/2}$, and $a_1$\,$=$\,$0.9$, we find a maximum observer angle 
$\theta_\gamma$\,$\sim$\,$0.043 \left( \theta_{\rm j} / 10^{-2} \right) \left( E / 10^{55} \textrm{ erg} \right)^{-0.74} \left(\delta t / 0.1 \textrm{ s} \right)^{1.5} \left( \Gamma_\textrm{tip} / 750 \right)^{3.7}$ rad, assuming the tip (i.e., the region of the jet at $\theta$\,$<$\,$\theta_\gamma$) is sufficiently fast in the first place. Thus, even if shallow power-law jets with narrow cores were intrinsically equally likely as `typical' top-hat jets with $\theta_{\rm j}$\,$\sim$\,$0.1$ rad, this already renders them about five times as rare. The intrinsic likelihood of producing a jet with a narrow core and shallow power-law structure further impacts their expected rate, as do the energetics and initial baryon loading of the jet.

\noindent \textbf{Observations and Data Analysis} 

\noindent \textbf{\textit{X-ray Observations}} 

\noindent \textit{Swift} 

The \textit{Neil Gehrels Swift Observatory} X-ray Telescope (XRT) began settled observations of GRB 221009A at 14:13:30 UT on 2022-10-09, approximately 143 s after the \textit{Swift}/BAT trigger and 3198 s after the \textit{Fermi}/GBM trigger. The afterglow was clearly detected and localized to RA, DEC (J2000) = $19^{h}13^m 03^{s}.85$, $+\ang{19;46;22.7}$ with an accuracy of \ang{;;3.5} (90$\%$ confidence level) \cite{Evans2009}. 
We obtained the XRT lightcurve from the \textit{Swift}-XRT GRB Lightcurve Repository\footnote{\url{https://www.swift.ac.uk/xrt_curves/}}. The lightcurve has been shifted in time by 3198 s to apply the \textit{Fermi}/GBM trigger as $T_0$. 
We further used the XRT Repository tools to create time-sliced spectra, in both windowed timing (WT) and photon counting (PC) modes, at multiple epochs for use in modeling the broadband spectral energy distribution of the GRB afterglow. 

\noindent \textit{XMM-Newton} 

We observed GRB 221009A using \textit{XMM-Newton} on November 10, 2022 through Directory's Discretionary Time (DDT; PI: Troja), corresponding to 32.72 d after the GRB trigger. The observations were performed in full frame mode using the thin filter, and lasted for 100 ks with a mid-time of 32.72 d. The data were reduced and analyzed using tasks with the Science Analysis System (\texttt{SAS v18.0.0}) software. We used a circular region of $60\arcsec$ to extract source photons. The background photons were taken from a nearby source-free circular region of $60\arcsec$.  We grouped the spectrum to a minimum of 1 count per bin and obtained the response matrix and ancillary response files using the \texttt{rmfgen} and \texttt{arfgen} tasks. 

\noindent \textit{NuSTAR} 

\textit{NuSTAR} observed GRB 221009A on five epochs at mid-times of 1.82, 5.92, 10.72, 23.93, and 31.95 d after the \textit{Fermi} trigger (see Table \ref{tab:Xobservations}). These data were obtained under two DDT programs (co-PIs: Margutti and Racusin; PI: Troja). The initial observations were first reported by \cite{GCN32695,GCN32788}. \textit{NuSTAR} is comprised of two identical focal plane modules, FPMA and FPMB, covering the $3-79$ keV energy range. The GRB afterglow is detected in each epoch. We reduced the data using the \textit{NuSTAR} Data Analysis Software pipeline (\texttt{NuSTARDAS}) within \texttt{HEASoft v6.29c}. The data were processed using \texttt{nupipeline}, and then lightcurves and spectra were extracted using \texttt{nuproducts}. In the initial observation at 1.8 d, source spectra were extracted from a 100\arcsec{} radius region centered on the transient. The background was similarly extracted from a 100\arcsec{} radius source-free region. For each of the latter observations, we used a 60\arcsec{} radius for both source and background regions.

\noindent \textbf{\textit{Optical/Near-Infrared Observations}}

\noindent \textit{MASTER}

MASTER is a system of wide-field fully robotic 40-cm telescopes with identical scientific equipment and real-time auto-detection at every observatory \cite{Lipunov2004,Lipunov2010,Kornilov2012}. The MASTER telescopes are MASTER-Amur, -Tunka, -Ural, -Kislovodsk, -Tavrida (Russia), -SAAO (South Africa), -IAC (Spain), -OAFA (Argentina). The MASTER Global Robotic Net \cite{Lipunov2010} began observing GRB 221009A using MASTER-SAAO on 2022-10-09 at 17:33:54 UT, corresponding to 0.18 d after the \textit{Fermi} trigger. The MASTER auto-detection system discovered the optical counterpart, MASTER OT J191303.43+194623.1 \cite{GCN32634}, as a transient in the initial image. The observations were obtained in the \textit{clear} ($C$) filter. Further observations were obtained with MASTER-OAF starting on 2022-10-09 at 23:56:38 UT, corresponding to 0.46 d after the trigger (see Table \ref{tab:MASTERobservations}). Aperture photometry was calibrated using 20 reference stars, and we applied a further correction to convert to the AB magnitude system. We note that earlier observations by MASTER-Tunka and MASTER-Amur, beginning at 93 s after the \textit{Fermi} trigger (2022-10-09 14:11:50 UT), were severely impacted by clouds and were unable to detect the source.

\noindent \textit{COATLI}

We observed the field of GRB 221009A with the 50-cm COATLI telescope at the Observatorio Astronómico Nacional on the Sierra de San Pedro Mártir in Baja California, México, and the HUITZI f/8 imager. We observed in $griz$ filters similar to Pan-STARRS1 (hereafter PS1; \cite{Chambers2016}) and took multiple 30 second images each night. The images were reduced using a custom pipeline, and then stacked to create a final image. Photometry was calculated for each stack using PSF fitting and calibrated relative to the PS1 catalog using all available field stars.

\noindent \textit{Gemini}

Target of Opportunity (ToO) observations of GRB 221009A were performed with the 8.1-m Gemini South Telescope in Cerro Pachon, Chile using GMOS-S and Flamingos-2 (hereafter F2) under two DDT programs (GS-2022B-DD-104, PI: O'Connor; GS-2022B-DD-103, PI: Rastinejad). The observations were first reported by \cite{GCN32749,GCN32750,GCN32860}.
Data were obtained at 4.4, 17.4 and 25.4 d. The observations were carried out in the $i$ filter with GMOS-S, and $J$, $H$, and $K_s$ filters with F2 (see Table \ref{tab:Optobservations}). We reduced and analyzed these data using the \texttt{DRAGONS}\footnote{\url{https://dragons.readthedocs.io/}} software \cite{Labrie2019}. The afterglow is clearly detected in each observation. 
We performed PSF photometry using \texttt{SExtractor} \cite{Bertin1996} and \texttt{PSFEx} \cite{Bertin2013}. The photometry was calibrated using nearby point-sources in the Two Micron All Sky Survey (2MASS) catalog \cite{Skrutskie2006}. We then applied a standard conversion between the Vega and AB magnitude systems.

\noindent \textit{Lowell Discovery Telescope} 

We obtained ToO observations with the Large Monolithic Imager (LMI) mounted on the 4.3-meter LDT at the Lowell Observatory in Happy Jack, AZ starting at 3.6 d after the GRB and continuing observations for 8 weeks under multiple programs (program PIs: Cenko, Hammerstein, O'Connor), see Table \ref{tab:Optobservations}. We first reported the initial observations in \cite{GCN32739,GCN32799}. The data were reduced with a custom pipeline \cite{Toy2016} that makes use of standard CCD reduction techniques in the IRAF\footnote{IRAF is distributed by the National Optical Astronomy Observatory, which is operated by the Association of Universities for Research in Astronomy (AURA) under cooperative agreement with the National Science Foundation (NSF).} package including bias subtraction, flat-fielding, and cosmic ray rejection \cite{vanDokkum2001}. Photometry was performed in the same manner as for Gemini, and calibrated against stars in the PS1 catalog. 

\noindent \textit{Palomar 200-inch}

We obtained data with the Palomar 200-in Wide InfraRed Camera (WIRC) in the $J$, $H$, and $K_s$ bands starting at 18:26:50 UT on 2022-11-02, taking 9 dithered images in each band for a total exposure of 495 s in $J$ and 330 s in $H$ and $K_s$ (Table \ref{tab:Optobservations}). The data were reduced using a \texttt{Python}-based reduction pipeline that uses standard image reduction techniques \cite{De2020}. The dithered images in each filter were astrometrically aligned and stacked to create a single composite image. The pipeline performed photometric calibration using the 2MASS point source catalog.

\noindent \textit{Other Optical and Near-Infrared Data}

We analyzed publicly available \textit{HST} observations (PI: Levan) obtained in WFC3/UVIS $F625W$ and $F775W$ filters and WFC3/IR $F098M$, $F125W$, and $F160W$ filters. The data were reduced and analyzed as detailed in \cite{OConnor2022}. 

We further supplemented our dataset with optical and infrared observations reported in GCN circulars \cite{GCN.32769, GCN.32818, GCN.32684, GCN.32743, GCN.32752, GCN.32652, GCN.32755, GCN.32646, GCN.32804, GCN.32678, GCN.32811, GCN.32758, GCN.32803, GCN32765hostz, GCN.32670,GCN.32662, GCN.32852, GCN.32795, GCN.32809, GCN.32753,GCN.32771, GCN.32756, GCN.32709}. 


\noindent \textit{\textbf{Radio Observations}}

The Australia Telescope Compact Array (ATCA) observed GRB 221009A as part of the DDT project CX515 on 2022-10-15, 2022-10-24, 2022-11-04, 2022-11-19, 2022-12-28, and 2023-01-27 (Table \ref{tab:Radioobservations}). The initial observation at 5.8 d after the GRB was first reported by \cite{GCN32763,GCN32766}. In the first two epochs, the three pairs of Intermediate Frequencies (IFs) were used 16.7/21.2 GHz, 33/35 GHz, 45/47 GHz and in the next two only 16.7/21.2 GHz and 33/35 GHz were used. In the last two epochs, corresponding to 80 and 110 d after the GRB, observations were obtained at 5.5 and 9 GHz. In all observing runs the array operated in continuum mode with the Compact Array Broadband Backend using 2048 channels over 2-GHz bandwidth per IF. The radio sources used as bandpass, primary and phase calibrator were 1921-293, 1934-638 and 1923+210, respectively. The phase calibrator was also used for reference pointing checks every hour. Standard procedures in the data reduction software package \texttt{MIRIAD}
\cite{Sault1995_MIRIAD}
were used to flag, calibrate the complex gains and image the phase calibrator and target. A robustness parameter value of two was used, and a 6-km baseline antenna was deselected during imaging.

We further used the \texttt{MIRIAD} task \texttt{calred} on the phase calibrator 1923+210 to test for phase decorrelation due to atmospheric turbulence. The task \texttt{calred} estimates the triple product amplitude, and, therefore, is unaffected by atmospheric turbulence induced phase decorrelation effects. A comparison between triple product and map flux density can be used to estimate the decorrelation factor. We found no significant decorrelation in the flux densities measured on the I-stokes restored maps, apart from three frequencies (35, 45 and 47 GHz) in Epoch 2 (2022-10-24; 14.7 d post-burst) at a level of 19\%, 14\% and 12\% respectively and on Epoch 4 (2022-11-19; 41 d post-burst) at 33 and 35 GHz of 4\% and 3\%, respectively. 
Seeing monitor data show a maximum rms path length of 
$\approx$500 microns during our second epoch (14.7 d) which, for a calibration interval of 15 min, is expected to result in a 25-50\% amplitude decorrelation above 30 GHz. Due to the large uncertainties, we exclude this epoch from our spectral analysis. 
No significant decorrelation is expected for observations at frequencies below 10 GHz (e.g., 80 and 110 d). 

The resulting flux densities and 5$\sigma$ upper limits are presented in Table \ref{tab:Radioobservations}. The errors are derived from summing in quadrature the statistical and systematic uncertainty (map rms and 5\% of the total flux as residual of gain calibration, respectively).

We supplemented our radio observations with data reported in GCN circulars  \cite{GCN32653,GCN32655,GCN32949,GCN32676}.

\newpage

\newpage

\bibliography{scibib}
\bibliographystyle{science}

\newpage

\newpage
\begin{table}[h]
\caption{Historical bright GRBs.} 
\label{tab:brightgrb}
\centering
\begin{tabular}{lcccc}
\hline\hline
\textbf{GRB} & \textbf{Observatory} & \textbf{Fluence} &  \textbf{Energy Band}   &  \textbf{Reference} \\
  &   & \textbf{(erg cm$^{-2}$)} &    &   \\ 
\hline 
840304 & PVO, ICE, Vela 5B & $2.8\times10^{-3}$ & 5 keV $-$ 3 MeV & \cite{Klebesadel84} \\
130427A & Fermi, Swift &  $2.5\times10^{-3}$ & 10 keV $-$ 1 MeV & \cite{Fermi130427A} \\
830801B & SIGNE 2 MP9 & $2.0\times10^{-3}$ & 30 keV $-$ 7.5 MeV &  \cite{Kuznetsov1986} \\
850624 & PVO & $2.0\times10^{-3}$ & 5 keV $-$ 3 MeV & \cite{Chuang90} \\
940703A & BATSE, GRANAT, ULYSSES & $1.6\times10^{-3}$ & 100 keV $-$ 1 MeV  & \cite{Tkachenko98} \\
\hline\hline
\end{tabular}
\end{table}

\begin{table}[h]
\caption{Log of X-ray observations of GRB 221009A analyzed in this work.} 
\label{tab:Xobservations}
\centering
\begin{tabular}{lccccc}
\hline\hline
\textbf{Start Time (UT)} & \textbf{Time (d)} & \textbf{Telescope} &   \textbf{Exp. (s)} & \textbf{ObsID} \\
\hline
2022-10-11 03:11:09	 & 1.82  & \textit{NuSTAR}  & 20665& 90802329002 \\
2022-10-15 05:21:09	 & 5.92  & \textit{NuSTAR} & 22465 & 90802329004 \\
2022-10-20 01:05:07	 & 10.72 & \textit{NuSTAR}   & 20444 &90802329006 \\
2022-11-02 06:05:07	 & 23.93 & \textit{NuSTAR}   & 21296 &90802329008 \\
2022-11-09 23:06:09 &  31.95 & \textit{NuSTAR} & 30022 & 90801331002  \\
2022-11-10 16:09:14 &  32.72 & \textit{XMM-Newton} &   103900 & 0913991901\\
\hline\hline
\end{tabular}
\end{table}

\begin{table}[h]
\caption{Log of optical and infrared observations of GRB 221009A analyzed in this work. } 
\label{tab:Optobservations}
\centering
\begin{tabular}{lcccccc}
\hline\hline
 \textbf{Time (d)} & \textbf{Telescope} & \textbf{Instrument} & \textbf{Filter} & \textbf{Exposure (s)} & \textbf{AB mag} \\
\hline
 3.6 &  LDT & LMI & $g$ & 1050 & $22.06\pm0.04$   \\
 3.6 &   LDT & LMI & $r$ & 600 & $20.44\pm0.04$  \\
3.6 &   LDT & LMI & $i$ & 400 & $19.37\pm0.01$  \\
3.6 &   LDT & LMI & $z$ & 300 & $18.67\pm0.01$ &  \\
4.4 & Gemini  & GMOS-S  & $i$ & 420 &  $19.78\pm0.02$  \\
4.4  &  Gemini & F2  & $J$ &  60 & $17.93\pm0.03$  \\
4.4   & Gemini  & F2 & $H$ & 60 & $17.23\pm0.03$   \\
 4.4   & Gemini  & F2  & $K_s$ & 60 & $16.69\pm0.02$  \\
 9.5    &  LDT &  LMI& $r$ & 600 & $21.85\pm0.08$ \\
 9.5     &   LDT &LMI   & $i$ & 450 & $20.75\pm0.05$  \\
 9.5 &   LDT  & LMI& $z$ & 300 & $20.01\pm0.03$   \\
 17.4 & Gemini  & GMOS-S &$i$ & 420  & $21.71\pm0.05$   \\
 17.4  &  Gemini & F2 & $J$  & 75  & $20.10\pm0.10$  \\
 17.4   & Gemini  & F2 & $H$ & 75  & $19.43\pm0.15$ \\
 17.4    & Gemini  &F2 &  $K_s$ & 75  & $18.94\pm0.08$    \\
  18.5   &  LDT &  LMI& $g$& 2400 & $24.71\pm0.15$ \\
  18.5       & LDT  & LMI & $r$ & 600  & $22.71\pm0.06$    \\
  18.5      & LDT  &  LMI& $i$ &  450 & $21.83\pm0.15$    \\
  18.5  &  LDT & LMI & $z$ &  300 & $20.97\pm0.05$    \\
  21.5  &  LDT & LMI & $r$ & 1350 & $22.91\pm0.06$   \\
  21.5  &  LDT & LMI & $i$ & 1200 & $21.88\pm0.05$   \\
  21.5    &  LDT & LMI & $z$ & 700 & $21.37\pm0.04$   \\
   24.2      & P200  &WIRC  &$J$ & 495 & $20.6\pm0.2$   \\
   24.2     & P200  &WIRC  & $H$ & 330 & $19.84\pm0.15$   \\
  24.2  &  P200 & WIRC & $K_s$ &330 & $19.48\pm0.15$   \\
  25.4  & Gemini  & F2 & $H$ & 245 & $19.95\pm0.07$    \\
   25.4 & Gemini  & F2 & $K_s$ & 225 & $19.54\pm0.07$    \\   
 28.5   & LDT  & LMI  & $r$ & 1200 & $23.55\pm0.15$    \\   
  28.5  & LDT  & LMI  & $i$ & 1050 &  $22.07\pm0.09$   \\   
 28.5   & LDT  & LMI  & $z$ & 1000 & $21.45\pm0.07$   \\   
  29.7   & \textit{HST}  & WFC3/UVIS & $F625W$ & 960 & $23.62\pm0.05$  \\   
  29.7  & \textit{HST}  & WFC3/UVIS & $F775W$ & 750 & $22.41\pm0.04$  \\   
  29.8  & \textit{HST}  & WFC3/IR & $F098M$ & 898 &  $21.24\pm0.01$  \\   
  29.8  & \textit{HST}  & WFC3/IR & $F125W$ & 798 &  $20.65\pm0.01$\\   
  29.8  & \textit{HST}  & WFC3/IR & $F160W$ & 798 &  $20.39\pm0.01$    \\   
  36.5  & LDT  & LMI & $r$ &450 & $23.64\pm0.11$   \\   
  36.5  & LDT  & LMI & $i$ &450 & $22.45\pm0.05$   \\   
  36.5  & LDT  & LMI & $z$ & 200 & $21.82\pm0.07$    \\   
\hline\hline
\end{tabular}
\end{table}

\begin{table}[h]
\centering
\begin{tabular}{lcccccc}
\hline\hline
 \textbf{Time (d)} & \textbf{Telescope} & \textbf{Instrument} & \textbf{Filter} & \textbf{Exposure (s)} & \textbf{AB mag} \\
\hline
  40.6  &  \textit{HST}  & WFC3/UVIS & $F625W$ & 960 & $23.92\pm0.06$   \\   
  40.6  &  \textit{HST}  & WFC3/UVIS & $F775W$ & 750 & $22.76\pm0.04$   \\ 
  52.5  & LDT  & LMI & $i$ & 1200 & $22.91\pm0.18$ &  \\
  55.7  &  \textit{HST}  & WFC3/UVIS & $F625W$ & 3775 & $24.38\pm0.05$    \\  
  56.3  &   \textit{HST} & WFC3/IR & $F098M$ & 898 &  $22.09\pm0.02$    \\   
   56.3 &   \textit{HST} & WFC3/IR & $F125W$ & 698 & $21.52\pm0.02$    \\  
  56.3  &   \textit{HST} & WFC3/IR & $F160W$ & 698 & $21.22\pm0.02$   \\ 
\hline\hline
\end{tabular}
\end{table}

\begin{table}[h]
\caption{Log of MASTER observations of GRB 221009A analyzed in this work. } 
\label{tab:MASTERobservations}
\centering
\begin{tabular}{lccccccc}
\hline\hline
 \textbf{JD} & \textbf{Time (d)} & \textbf{Instrument} & \textbf{Filter} & \textbf{Exposure (s)} & \textbf{AB mag}  \\
\hline
2459862.23 & 0.17945982 & MASTER-SAAO & $C$ & 180  & $15.47\pm0.03$ \\
2459862.24 & 0.1819111  & MASTER-SAAO & $C$ & 180  & $15.73\pm0.01$ \\
2459862.25 & 0.19427623 & MASTER-SAAO & $C$ & 180  & $15.71\pm0.09$ \\
2459862.25 & 0.19672313 & MASTER-SAAO & $C$ & 180  & $15.8 \pm0.02$ \\
2459862.25 & 0.19917371 & MASTER-SAAO & $C$ & 180  & $15.89\pm0.06$ \\
2459862.26 & 0.20162035 & MASTER-SAAO & $C$ & 180  & $15.85\pm0.03$ \\
2459862.26 & 0.20407257 & MASTER-SAAO & $C$ & 180  & $15.76\pm0.03$ \\
2459862.26 & 0.20638141 & MASTER-SAAO & $C$ & 180  & $15.86\pm0.04$ \\
2459862.26 & 0.20813872 & MASTER-SAAO & $C$ & 180  & $15.87\pm0.05$ \\
2459862.26 & 0.20919847 & MASTER-SAAO & $C$ & 180  & $15.76\pm0.04$ \\
2459862.26 & 0.21011946 & MASTER-SAAO & $C$ & 180  & $15.69\pm0.07$ \\
2459862.27 & 0.2111838  & MASTER-SAAO & $C$ & 180  & $16.00\pm0.06$ \\
2459862.27 & 0.21293514 & MASTER-SAAO & $C$ & 180  & $15.81\pm0.03$ \\
2459862.27 & 0.21524762 & MASTER-SAAO & $C$ & 180  & $15.97\pm0.06$ \\
2459862.27 & 0.21769996 & MASTER-SAAO & $C$ & 180  & $15.92\pm0.08$ \\
2459862.28 & 0.22258312 & MASTER-SAAO & $C$ & 180  & $15.95\pm0.07$ \\
2459862.3  & 0.24573242 & MASTER-SAAO & $C$ & 360  & $15.91\pm0.07$ \\
2459862.3  & 0.24781575 & MASTER-SAAO & $C$ & 360  & $16.13\pm0.18$ \\
2459862.32 & 0.26250605 & MASTER-SAAO & $C$ & 360  & $16.07\pm0.12$ \\
2459862.32 & 0.26739103 & MASTER-SAAO & $C$ & 360  & $15.99\pm0.14$ \\
2459862.33 & 0.27228741 & MASTER-SAAO & $C$ & 540  & $16.00\pm0.13$ \\
2459862.51 & 0.46092935 & MASTER-OAFA & $C$ & 900  & $16.38\pm0.02$ \\
2459862.53 & 0.47889935 & MASTER-OAFA & $C$ & 900  & $16.43\pm0.02$ \\
2459862.56 & 0.50871935 & MASTER-OAFA & $C$ & 1800 & $16.5 \pm0.03$ \\
2459862.59 & 0.54036935 & MASTER-OAFA & $C$ & 1980 & $16.43\pm0.03$ \\
2459863.55 & 1.49341128 & MASTER-OAFA & $C$ & 1980 & $18.11\pm0.16$ \\
\hline\hline
\end{tabular}
\end{table}

\begin{table}[h]
\caption{Log of COATLI observations of GRB 221009A analyzed in this work.}
\label{tab:COATLIobservations}
\centering
\begin{tabular}{lccc}
\hline\hline
 \textbf{Time (d)} & \textbf{Filter} & \textbf{Exposure (s)} & \textbf{AB mag}  \\
\hline
1.71 &	$r$ &	405 &	$19.58 \pm	0.135$ \\
1.71 &	$z$ &	405 &	$18.05 \pm	0.09$ \\
1.71 &	$i$ &	405 &	$18.38 \pm	0.06$ \\
2.66 &	$i$ &	7560 &	$18.93 \pm	0.02$ \\
3.63 &	$i$ &	5670 &	$19.38 \pm	0.03$ \\
3.64 &	$z$ &	5670 &	$18.95 \pm	0.05$ \\
4.65 &	$z$ &	4320 &	$19.40 \pm	0.07$ \\
4.65 &	$i$ &	4320 &	$19.72 \pm	0.04$ \\
5.62 &	$z$ &	2580 &	$19.03 \pm	0.12$ \\
5.65 &	$i$ &	6360 &	$20.11 \pm	0.07$ \\
6.64 &	$z$ &	3555 &	$19.84 \pm	0.12$ \\
6.64 &	$i$ &	3570 &	$20.27 \pm	0.07$ \\
8.69 &	$i$ &	2160 &	$20.62 \pm	0.18$ \\
10.67 &	$i$ &	3240 &	$20.95 \pm	0.15$ \\
11.63 &	$i$ &	8730 &	$21.22 \pm	0.10 $\\
14.61 &	$i$ &	11280 &	$21.31 \pm	0.11$ \\
15.61 &	$i$ &	4425 &	$21.02 \pm	0.13$ \\
18.61 &	$i$ &	9165 &	$21.45 \pm	0.15$ \\
19.61 &	$i$ &	2700 &	$21.92 \pm	0.28$ \\
20.61 &	$i$ &	9180 &	$21.85 \pm	0.18$ \\
21.60 &	$i$ &	9180 &	$21.63 \pm	0.21$ \\
27.58 &	$i$ &	8100 &	$22.21 \pm	0.29$ \\
\hline\hline
\end{tabular}
\end{table}

\begin{table}[h]
\caption{Log of radio observations of GRB 221009A analyzed in this work.}
\label{tab:Radioobservations}
\centering
\begin{tabular}{lcccc}
\hline\hline
\textbf{Start Time (UT)} & \textbf{Time (d)} & \textbf{Telescope} & \textbf{$\lambda_0$ (GHz)} & \textbf{Flux density (mJy)}  \\
\hline
2022-10-15 05:03:54  & 5.78 & ATCA  & 16.7 & $6.7\pm0.3$  \\ 
2022-10-15 05:03:54  & 5.78 & ATCA  & 21.2 & $5.8\pm0.3$ \\ 
2022-10-15 05:03:54  & 5.78 & ATCA  & 33 & $4.6\pm0.2$  \\ 
2022-10-15 05:03:54 & 5.78 & ATCA  & 35 & $4.5\pm0.2$  \\ 
2022-10-15 05:03:54  &5.78  & ATCA & 45 & $3.5\pm0.2$  \\ 
2022-10-15 05:03:54  & 5.78 & ATCA & 47 &  $3.1\pm0.2$ \\ 
2022-10-24 03:07:44  & 14.72 & ATCA  & 16.7 & $2.16\pm0.11$   \\ 
2022-10-24 03:07:44  & 14.72 & ATCA  & 21.2 & $2.32\pm0.17$    \\ 
2022-10-24 03:07:44  & 14.72 & ATCA & 33 & $1.12\pm0.13$   \\ 
2022-10-24 03:07:44 & 14.72 & ATCA& 35 & $0.91\pm0.14$   \\ 
2022-10-24 03:07:44  & 14.72 & ATCA & 45 & $0.68\pm0.08$    \\ 
2022-10-24 03:07:44  & 14.72 & ATCA & 47 &  $0.22\pm0.06$  \\ 
2022-11-04 01:25:15 &  25.68 & ATCA  & 16.7  & $2.00\pm0.10$   \\ 
2022-11-04 01:25:15  & 25.68  & ATCA  & 21.2  & $1.85\pm0.10$    \\ 
2022-11-04 01:25:15  & 25.68  & ATCA  & 33  & $1.57
\pm0.09$   \\ 
2022-11-04 01:25:15 &  25.68 & ATCA  & 35  & $1.32\pm0.08$   \\ 
2022-11-19 06:25:10 &  40.71 & ATCA  & 16.7  &  $1.32\pm0.08$  \\ 
2022-11-19 06:25:10  & 40.71  & ATCA  & 21.2  & $0.72\pm0.14$    \\ 
2022-11-19 06:25:10  & 40.71  & ATCA  & 33  & $<0.45$   \\ 
2022-11-19 06:25:10 &  40.71 & ATCA  & 35  & $<0.60$   \\ 
2022-12-28 01:14:15 & 79.60 & ATCA  & 5.5 & $0.82\pm0.05$    \\ 
2022-12-28 01:14:15 & 79.60 & ATCA  & 9 & $0.42\pm0.03$    \\ 
2023-01-27 00:09:15 & 109.53 & ATCA & 5.5 & $0.49\pm0.04$   \\ 
\hline\hline
\end{tabular}
\end{table}

\end{document}